\documentclass[manuscript]{aastex}

\usepackage{color}

\slugcomment{Not to appear in Nonlearned J., 45.}

\shorttitle{X-ray Spectral Variability of the Radio Galaxy Centaurus A}
\shortauthors{Fukazawa et al.}

\begin{document}


\title{Suzaku view of X-ray Spectral Variability of the Radio Galaxy
Centaurus A : Partial Covering Absorber, Reflector, and 
Possible Jet Component}


\author{Yasushi Fukazawa\altaffilmark{1},
Kazuyoshi Hiragi\altaffilmark{1}, 
Syoko Yamazaki\altaffilmark{1}, \\
Motohiro Mizuno\altaffilmark{1}, 
Kazuma Hayashi\altaffilmark{1}, 
Katsuhiro Hayashi\altaffilmark{1}, 
Sho Nishino\altaffilmark{1}, 
Hiromitsu Takahashi\altaffilmark{1}, 
and Masanori Ohno\altaffilmark{1}}
\affil{Department of Physical Science, Hiroshima University, 1-3-1
Kagamiyama, Higashi-Hiroshima, Hiroshima 739-8526, Japan}
\email{fukazawa@hep01.hepl.hiroshima-u.ac.jp}



\begin{abstract}
We observed a nearby radio galaxy, the Centaurus A (Cen A), three times 
with Suzaku in 2009,
and measured the wide-band X-ray spectral variability more accurately 
than the previous measurements.
The Cen A was in the active phase in 2009, and the flux became higher by
a factor of 1.5--2.0 and the spectrum became harder than that in 2005.
The Fe-K line intensity increased by 20--30\% from 2005 to 2009.
The correlation of the count rate between the XIS 3--8 keV and PIN
15--40 keV band showed a complex behavior with a deviation from a
linear relation.
The wide-band X-ray continuum in 2--200 keV can be fitted with an absorbed
powerlaw model plus a reflection component, 
or a powerlaw with a partial covering Compton-thick absorption.
The difference spectra between high and low flux periods in each
observation were reproduced by a powerlaw with a partial covering
Compton-thick absorption.
Such a Compton-thick partial covering absorber was 
for the first time observed for the Cen A.
The powerlaw photon index of the difference spectra in 2009 is almost
 the same as that of the time-averaged spectra in
 2005, but steeper by $\sim0.2$ than that of the time-averaged spectra in
 2009.
This suggests an additional hard powerlaw component with a photon
index of $<1.6$ in 2009.
This hard component 
\textcolor{black}{
could be
}
a lower part of the inverse-Compton-scattered component 
from the jet, whose gamma-ray emission 
has recently been detected with the Fermi/LAT.
\end{abstract}





\section{Introduction}

Radio galaxies host an extended emission from relativistic jets
and lobes.
Jet emission is extremely enhanced by relativistic effects for
blazars, whose jet direction is close to the line of sight, and
observed in the multi-wavelength band from radio to TeV gamma-rays.
Recently, Fermi gamma-ray space telescope has opened a new era for
studying jet emissions, by detecting more than 500 gamma-ray
blazars (Abdo et al. 2010a).
Furthermore, Fermi detected 10 radio galaxies and revealed that 
radio galaxies are also gamma-ray emitters (Abdo et al. 2010b).
Radio galaxies are important to study the jet structure from the
mis-aligned jet direction; jet emissions of blazars
are dominated by the central region of jets due to the beaming effect
while those of radio galaxies are weighted by the jet outer boundary.
\textcolor{black}{
Possible jet emission from radio galaxies has been
reported from infrared to X-ray band (Chiaberge et al. 1999; Hardcastle
et al. 2000; Hardcastle et al. 2006), but the spectral property is still
uncertain, especially for FR-I radio galaxies, due to the contribution
of the accretion disk.
}

Centaurus A (Cen A) is the nearest radio galaxy, 
and its gamma-ray emission has
been established by Fermi (Abdo et al. 2009a, 2010c) and 
HESS (Aharonian et al. 2009), 
and is the second brightest GeV gamma-ray radio galaxy,
following NGC 1275 (Abdo et al. 2009b, 2010b). 
The GeV gamma-ray emission does not come from the kiloparsec-scale jet but
likely from the beamed sub-arcsec jet resolved by VLBI (Abdo et
al. 2009c; Mueller et al. 2011).
In addition, Fermi found that sub-Mpc giant radio lobes of the 
Cen A are also GeV gamma-ray emitters (Abdo et al. 2010d).
\textcolor{black}{
Nonthermal soft X-ray emission from the sub-arcsec jet is suggested for
the Cen A (Evans et al. 2004).
}
On the other hand, Chandra resolved the X-ray emission from
kiloparsec-scale jets (Kraft et al. 2000).
Cen A is the brightest 
\textcolor{black}{
AGN
}
in the hard X-ray band (Tueller et al. 2008), and
the X-ray spectrum
is very similar to that of Seyfert galaxies (Wang et al. 1986,
Kinzer et al. 1995, Rothschild et al. 1999).
On the other hand, it seems to smoothly connect to the MeV/GeV emission
detected with CGRO COMPTEL/EGRET, like blazars (Steinle et al. 1998, 
Sreekumar et al. 1999).
Suzaku observed the Cen A in 2005, and it was reported that 
Seyfert-like X-ray emission was dominated (Markowitz et al. 2007).
Based on the INTEGRAL time-averaged spectra with a long exposure,
 Beckmann et al. (2011) also indicated that the origin of the hard
X-rays is preferred
to be a Seyfert-like emission but the nonthermal emission scenario
cannot be ruled out.
Another question on the Cen A X-ray spectrum is that the reflection
continuum was often not required in the spectral fitting (Rothschild et
al. 1999; Markowitz et al. 2007; Rothschild et al. 2011; Beckmann et
al. 2011), regardless of the existence of a neutral Fe-K fluorescence line.
Therefore, interpretation of the Cen A X-ray spectrum still has 
several open issues.

However, detailed studies of spectral variation 
\textcolor{black}{
in the hard X-ray band
}
have not been reported yet.
Time variability of X-ray spectra gives us important opportunities to decompose
the spectral components.
The Cen A spectrum is strongly absorbed in the soft X-ray band like
Seyfert 2 galaxies, and therefore hard X-ray study of time variation is
important.
In order to measure the spectral variability in detail, we again
observed the Cen A with Suzaku, which enables us to measure the
short-term time variation in the hard X-ray band with the best
accuracy.
In this paper, we report the studies of X-ray spectral variability
of the Cen A with Suzaku (Mitsuda et al. 2007).
Throughout this paper, we assumed the distance to the Cen A as 3.8 Mpc
(Rejkuba 2004), and the errors are shown as a 90\% confidence level.
We refer to the solar photospheric values  
(Anders \& Grevesse 1989) for the solar abundance ratio of the
photoelectric absorption, reflection, and plasma
model. 
The cross-section for absorption models is set to that of 
Baluci${\rm \acute{n}}$ska-Church \& McCammon (1992).

\section{Observation and Data Reduction}

\begin{figure}[htb]
\centering
\includegraphics[width=0.5\textwidth]{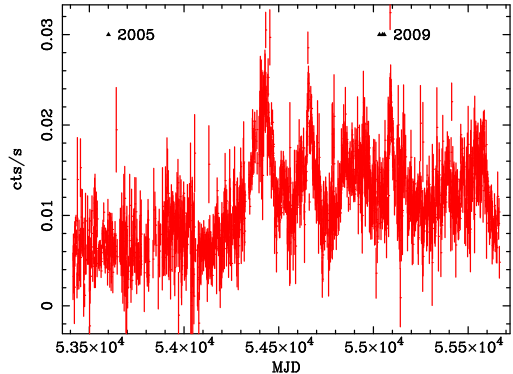}
\caption{Swift/BAT light curve of the Cen A 
\textcolor{black}{
in the 15--50 keV band
}
. Triangles represent the
 period of Suzaku observations.}
\label{batlc}
\end{figure}

We observed the Cen A three times with Suzaku on July 20--21, 
August 5--6, and
August 14--16 in 2009, as summarized in table \ref{obs}.
In this paper, we also analyzed the Suzaku data obtained in 2005 in the
same way for a comparison.
Fugure \ref{batlc} shows the periods of Suzaku observations on the
Swift/BAT light curve
\footnote{\tt
http://swift.gsfc.nasa.gov/docs/swift/results/transients/}.
The Cen A entered an active phase since 2007 summer.
The 2005 data correspond to the low state, while the 2009 data do to the
high state.
All the observations were performed with XIS 5x5 or 3x3 modes 
(Koyama et al. 2007) and 
normal HXD mode (Takahashi et al. 2007, Kokubun et al. 2007), 
except for the XIS observation in 2005, which was
operated in 5x5, 3x3, or 2x2 modes with 1/4 window option  
in order to avoid the pile-up.
The Cen A was observed at the XIS nominal position in 2005, while it was
observed at the HXD 
nominal position in 2009.
The XIS count rate was 7--10 counts sec$^{-1}$ in 2009, and thus the
pile-up did not occur.
We also confirmed no pile-up affections
by checking that the results of spectral fittings did not change
when we excluded the central 2-arcmin region.

We utilized the data processed with the Suzaku version 2.4 pipeline 
software, and
performed the standard data reduction with criteria such as 
a pointing difference of $<1.5^{\circ}$, a elevation angle of
$>5^{\circ}$ from the earth rim, a geomagnetic cut-off rigidity (COR) 
of $>$6 GV,
and the SAA-elapsed time of $>256$ s.
Further selection was applied with criteria such as an
earth elevation angle of $>20^{\circ}$ 
for the XIS, COR$>$8 GV and the SAA-elapsed time (T\_SAA\_HXD) of 
$>$500 s for the HXD.
XIS photon events were accumulated within 4 arcmin of the Cen A nucleus, with 
the XIS- 0, 2, and 3 data coadded.
The XIS rmf and arf files were created with {\tt xisrmfgen} and {\tt
xisarfgen} (Ishisaki et al. 2007), respectively, and
the XIS detector background was estimated with {\tt xisnxbgen} (Tawa et
al. 2008).
For the HXD, the "tuned" PIN and GSO background was used (Fukazawa
et al. 2009) and
the good time interval (GTI) was determined by taking the logical-and
of GTIs among the data and background model.
Since the XIS light-leakage estimation was not valid at the beginning of the
2005 observation, we eliminated the data in the first 12 hours.
Note that Markowitz et al. (2007) included this period in the analysis
and therefore our results are somewhat different from theirs for
absorption model parameters and so on.
For the XIS and HXD-PIN, 
CXB was added to the background spectrum thus obtained, although it is
negligible for the HXD-GSO.
The latest GSO response file (version 20100524) and 
the GSO response correction file (version 20100526) were utilized.
The former is updated in terms of energy-channel linearity and GSO gain
history (Yamada et al. 2011), and 
the latter compensated
for any disagreement of 10--20\% between the Crab spectral model 
and the data
\footnote{\tt http://www.astro.isas.jaxa.jp/suzaku/analysis/hxd/gsoarf2/}.


As a result,
a net exposure time for the XIS is around 33 ks, 62 ks, 51 ks, and 56 ks 
for the 2005, 2009-1st, 2nd, and 3rd observation, respectively.
The exposure time for the HXD is about 70\% of the XIS ones, due to 
additional cuts of high background periods.
The HXD-PIN signal rate is higher than the background rate below 50 keV,
while the HXD-GSO signal rate is $<10$\% of the background rate.
Therefore, we checked the reproducibility of the GSO background as
described in appendix.
As a result, the reproducibility of the GSO background is found to be as
good as around 1\%.
Note that 
the data of the 2009 1st observation was already used in Abdo et al. (2010c).

\section{Data Analysis}

\subsection{Light Curves and Correlations}

\begin{figure}[htb]
\centering
\includegraphics[width=0.5\textwidth]{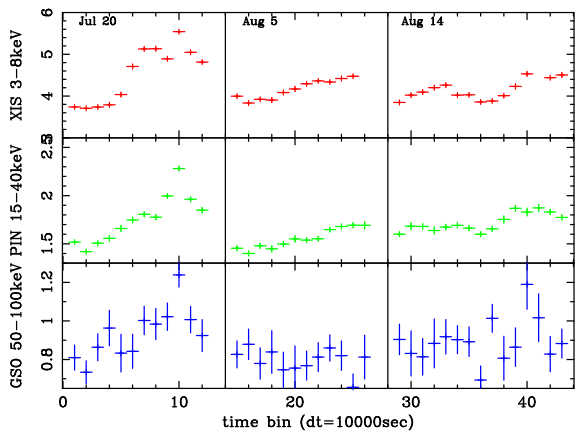}
\caption{Suzaku light curve of the Cen A for the 2009 Suzaku
 observations. 
From top to bottom, XIS-F (3--8
 keV), PIN (15--40 keV), and GSO (50--100 keV) are presented. The
 horizon axis indicates the time bin number with a step of 10000 sec.  
The beginning of light curves is 55032.328958, 55048.284699, 
55057.380162 in MJD, for the 1st, 2nd, and 3rd observation, respectively.}
\label{lc}
\end{figure}

Figure \ref{lc} shows the count-rate light curves of the XIS (3--8 keV),
PIN (15--40 keV), and GSO (50--100 keV), with a time bin of 10 ks.
The background was subtracted for the PIN and GSO, while the XIS
background rate is $<1$\% of the signal, and thus is negligible.
Note that the background rate of the PIN and GSO is around 0.3 and 8--10
count s$^{-1}$, respectively.
All the observations clearly exhibit time variability in the 
XIS and PIN light curves with an amplitude of up to 50\% and a time
scale of 10--20 ks.
This time scale is reasonable for the black hole mass of 
$(0.5-1)\sim10^8 M_{\odot}$ (Silge et al. 2005; Krajnovuie et al. 2007;
Neumayer et al. 2007).
The largest variability occurred during the 2009-1st observation.
Looking at the light curves, there is a different variability
pattern between the XIS and PIN.
For example, at 0--50000 sec (1--5th bin) in the 2009-1st observation, 
the rising trend of the count rate is linear for the PIN and concave for
the XIS.
At 0--70000 (29--35th bin) sec in the 2009-3rd observation, 
the variability pattern is also different between the XIS and PIN.
For the GSO light curve, the error is somewhat large, but a similar trend
of variability is clearly seen.

\begin{figure}[htb]
\centering
\includegraphics[width=0.5\textwidth]{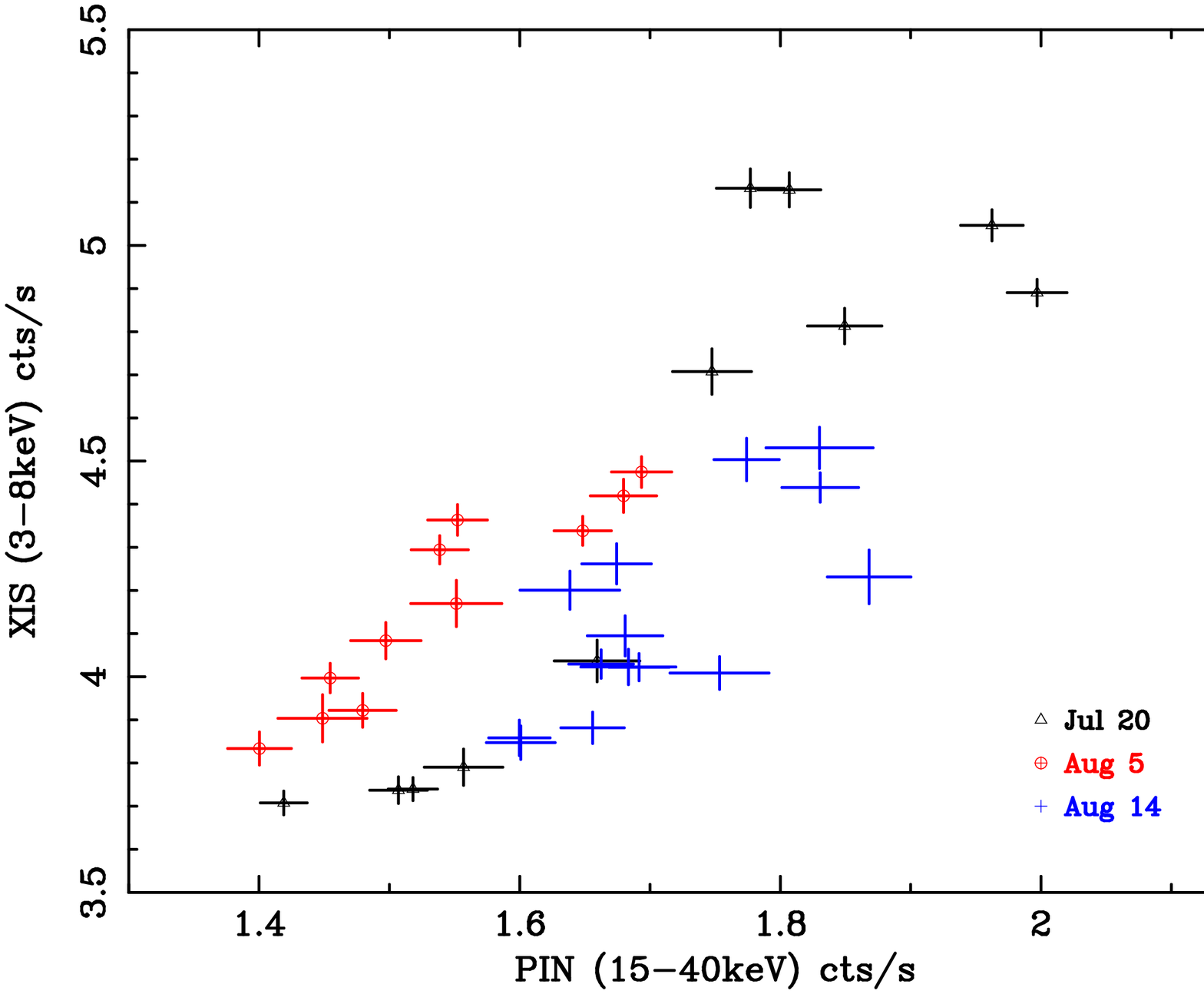}
\includegraphics[width=0.5\textwidth]{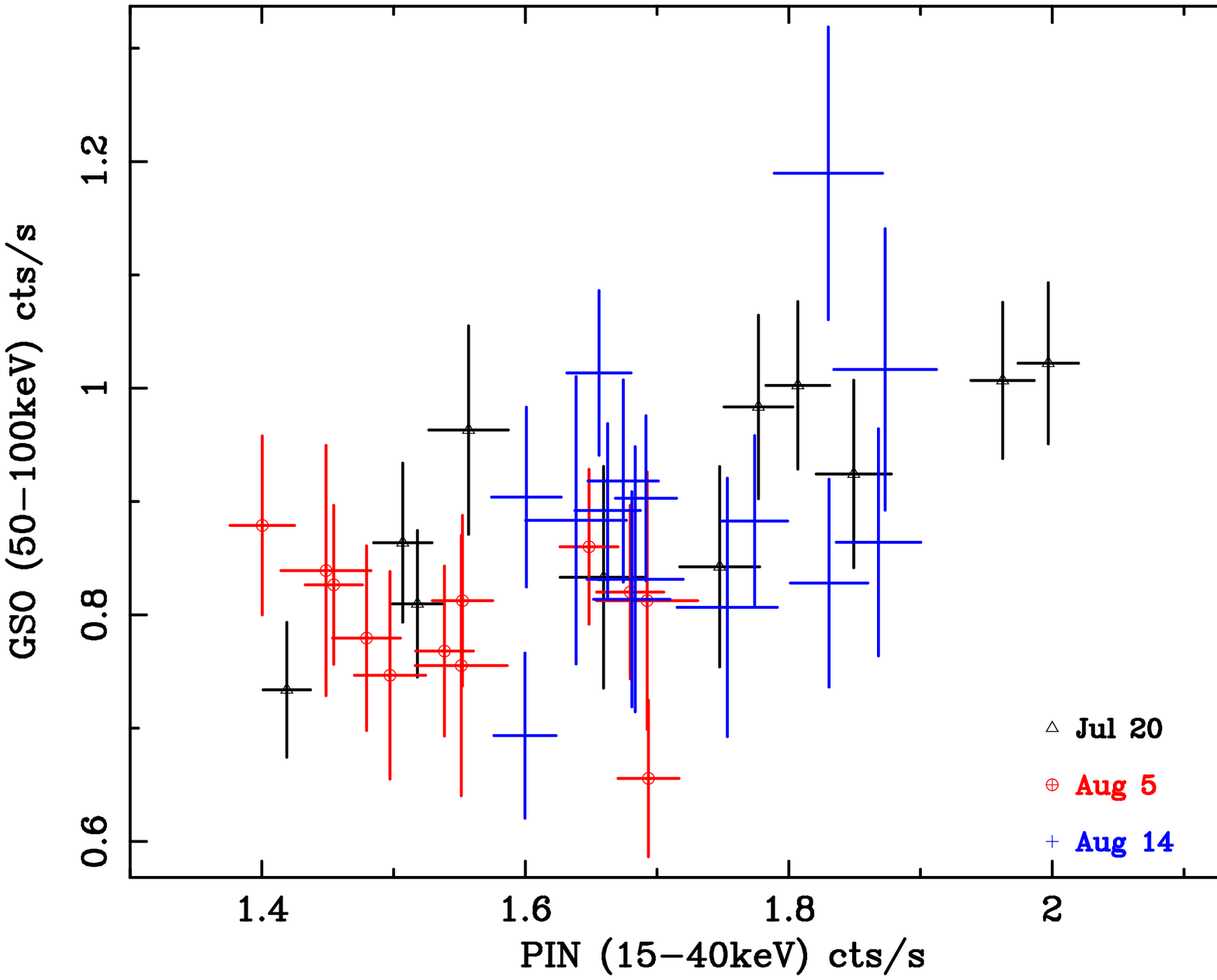}
\caption{Correlation of count rates of the XIS (3-8keV; left) and the
GSO (50-100keV; right) against the PIN (15-40keV). 
Triangles, circles, and crosses are the data of the 1st, 2nd, and 3rd
observation in 2009.}
\label{correlation}
\end{figure}

Figure \ref{correlation} shows a correlation of the count rate 
among different energy bands.
We do not treat the 2005 data, since the XIS to PIN effective area ratio
is different between 2005 and 2009 
and the data duration of the 2005 observation was short.
The XIS count rate in 3--8 keV generally correlates with the PIN count
rate in 15--40 keV, but the slope is different among observations.
Furthermore, the correlation is not completely linear,
especially for the 2009-1st and 3rd observation; the deviation is up to
20\% or so.
These trends indicate that the spectral shape varied significantly,
suggesting multiple spectral components or a change of the spectral shape.
On the other hand, considering that the GSO background reproducibility is
around 0.1--0.2 count s$^{-1}$ (Fukazawa et al. 2009),
it can be said that the GSO count rate in 50--100 keV
correlates with the PIN count rate within 10--20\%.
This situation is also the same as that in 100--200 keV.
Therefore, the emission in 15--200 keV could be mainly explained by a
combination of a
variable component and a constant one within errors.

\subsection{Modeling of the Soft X-ray Component}

The X-ray spectrum of the Cen A is known to consist of roughly two
components; a spatially extended component in the soft band and a
strongly absorbed hard nuclear component (Markowitz et al. 2007).
\textcolor{black}{
The former extended component was clearly resolved into many complex
features associated with the Cen A jets, such as kiloparsec-jets, shock
regions, together with interstellar hot medium and discrete sources in
the parent galaxy (Kraft et al. 2008).
Especially, shock regions show a relatively hard powerlaw emission
(Croston et al. 2009), whose flux is lower by two orders of magnitudes
than that of the nuclear X-ray emission. 
}
Since the spectral component in the soft band somewhat affects the
modeling of the nuclear component, we first modeled the soft component by
using the XIS data of the 2009 2nd observation.
Before this analysis, we fitted the XIS spectra in 2--10 keV, 
to model the hard X-ray continuum with the absorbed powerlaw model.
Then, we included the model of the hard component whose parameters are
fixed to the values obtained above, except for the powerlaw normalization, 
and fitted the XIS spectra in 0.7--3 keV with one {\tt apec} thermal plasma
model, together with the absorbed powerlaw model.
The metal abundance is left free, and the photoelectric absorption
model {\tt phabs} is multiplied.
The relative normalization between the XIS-F and XIS-B is let to be
free, 
\textcolor{black}{
due to calibration uncertainties associated with attitude fluctuations.
}
We ignored the 1.82--1.84 keV band due to the XIS calibration problems.
However, this modeling could not reproduce the spectra with a reduced
$\chi^2$ values of 4.14, since the {\tt apec}
model cannot simultaneously explain the emission lines and the
continuum around 1.5--2 keV where the spectral slope is around 2.
Next, we added the {\tt bremss} model with the absorption to model the
\textcolor{black}{
extended hard emission from unresolved point sources and jet features 
}
in the Cen A (Matsushita et al. 1994).
\textcolor{black}{
Many sources in the Cen A
}
could contribute to the X-ray 
emission, but the obtained
X-ray luminosity of the {\tt bremss} component is $8\times10^{39}$ erg
s$^{-1}$ (2--10 keV), which can be explained by the sum of X-ray point
sources in the Cen A 
\textcolor{black}{
(Kraft et al. 2001)
}
.
This is not the matter of this paper, 
and we do not further discuss about it in this paper.
This model made the fit improve, and a reduced $\chi^2$ value became 1.44, 
but the emission lines could not be well reproduced.
Markowitz et al. (2007) reported that two-temperature plasma
components were required.

\begin{figure}[htb]
\centering
\includegraphics[width=0.5\textwidth]{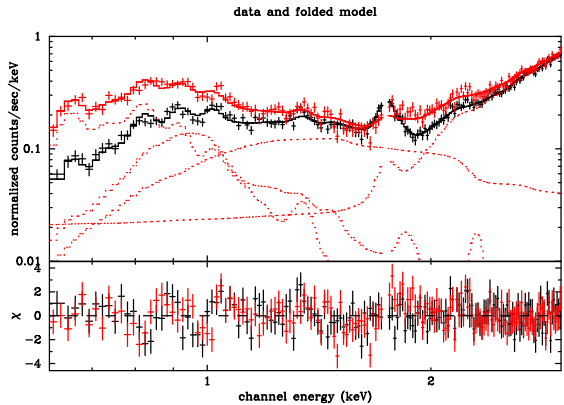}
\caption{Spectral fitting of the XIS spectra in 2009 2nd observation
 with the two-temperature {\tt apec} plasma models plus bremmstrahlung, together with the
 emission model of the nuclear emission seen above 2 keV. Details are
 described in the text. Solid line represents the best-fit total
 emission model, and dotted lines represent each of spectral component;
 two {\tt apec}, {\tt bremss}, and absorbed powerlaw. Red and black ones
 correspond to the XIS-B and XIS-F, respectively.}
\label{xisapec2}
\end{figure}

We thus added one more {\tt apec} model whose temperature and abundance are
left free.
Then, the spectrum is well fitted with the best-fit absorption column density
and metal abundance of 
$1.6\times10^{21}$ cm$^{-2}$ and $\sim0.3$ solar, respectively, 
but their errors became larger.
The former best-fit value is somewhat larger than that of the Galactic value 
$8.6\times10^{20}$ cm$^{-2}$ (Dickey \& Lockman 1990), implying an
additional intrinsic absorber such as 
\textcolor{black}{
interstellar medium within the parent galaxy
}.
We hereafter fixed the absorption column density and metal abundance to
the above values.
Evans et al. (2004) and 
Markowitz et al. (2007) reported the detection of Si and S fluorescence
lines.
In this analysis, the Si-K line is not significant with an upper limit
of $7\times10^{-6}$ c s$^{-1}$ cm$^{-2}$.
Therefore, we include the S-K line at 2.306 keV in the model and then
the fit improved with $\Delta\chi^2=10$, and its intensity is
$1.57_{-0.97}^{+0.70}\times10^{-5}$ c s$^{-1}$ cm$^{-2}$.
As a result, the soft X-ray spectra could be fitted well with a reduced
$\chi^2$ values of 1.25, as shown in
figure \ref{xisapec2} and 
the best-fit parameters are summarized in table \ref{xisapec}.

We also analyzed the XIS spectra of other observations in the same way
as above, and the results are summarized in table \ref{xisapec}.
Since the thermal components are believed to be extended, their parameters
should be constant and the best-fit values are consistent among all the 
observations within errors.
In the following broad-band fitting for all the observations, 
we fixed the temperatures and column densities of the
above model to the values in table \ref{xisapec}.

\begin{figure}[htb]
\centering
\includegraphics[width=0.5\textwidth]{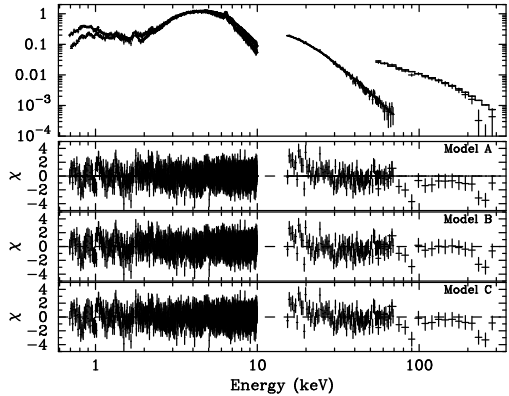}
\caption{Spectral fitting of the XIS/PIN/GSO spectra in 2009 2nd observation
 with the model A, B, and C. Detailed description of the models are 
described in the text. 
Top panel shows the XIS/PIN/GSO spectra with the best-fit model C. 2nd, 3rd, and 4th panels show the residual of the fitting for the model
 A, B, and C, respectively.}
\label{specabc}
\end{figure}

\subsection{Time-Averaged Spectrum of Each Observation}

Although the spectral variability during each observation is indicated
in the previous subsection, we here analyze the time-averaged spectrum
of each observation in order to understand the spectral shape.
At first, we describe the spectral fitting of the 2009-2nd observation,
whose time variability was relatively moderate.
After determining the best-fit modeling, we apply it to other
observations.

We then fit the XIS-F, 
XIS-B, PIN, and GSO spectra of the 2009-2nd observation in 0.7--300 keV 
simultaneously to model the nuclear hard component.
For each detector, the energy range of 0.7--10 keV, 15--70 keV, and
55--300 keV for the XIS, PIN, and GSO, respectively, are used in the
fitting.
A relative normalization between the XIS-F and XIS-B is left to be free,
while that between the XIS and PIN, or between the PIN and GSO, 
is fixed to 1.18 (Maeda et al. 2008) or 1.0
\footnote{\tt
http://www.astro.isas.jaxa.jp/suzaku/analysis/hxd/gsoarf2/}, respectively.
The spectral models for the soft X-ray emission were included, but
the model parameters of two {\tt apec} models and {\tt bremss} model
are fixed,
and normalizations of the higher-temperature {\tt apec} model
and the {\tt bremss} model are left free.

For the hard nuclear component, 
we at first tried the basic model; an absorbed power-law with a Fe-K line
(model A).
For the strong photoelectric absorption for the nuclear emission, we
use the {\tt zvphabs} model, where the Fe abundance is left free and
other elemental abundances are fixed to 1 solar.
We tentatively included the high energy cut-off fixed at 1000 keV for
the powerlaw model.
The broad-band X-ray spectra in 0.7--300 keV are overall fitted with
this basic model. 
The residual of fitting with this model is shown in the 1st panel of figure
\ref{specabc}, where a reduced $\chi^2$ value is 1.21.
However, significant residual structures are seen around 
15--30 keV and 50--300 keV.
The residual around 15--30 keV could be due to the reflection component,
and then we include the {\tt pexrav} model (Zdziarsk et al. 1995) 
for the reflection (model B),
where the input powerlaw parameters of the {\tt pexrav} model are 
tied to those of the 
direct nuclear powerlaw model and the inclination angle is assumed to be
$cos\theta =0.5$ ($\theta=60^{\circ}$); 
only the free parameter is the reflection fraction
$R$.
Furthermore, we multiplied the reflection component by the absorption.
As shown in the 2nd panel of figure \ref{specabc},
the residual around 15--40 keV still remains a little, but the fit
improved with $\Delta\chi^2=35$ for addition of two parameters.
The reflection fraction became $R=0.19$, while 
the absorption for the reflection component is not required.

Alternatively, the residual structure in the 1st panel of figure \ref{specabc} 
could be reproduced 
by a partial covering absorption
with a column density of $\sim10^{24}$ cm$^{-2}$.
Then, instead of the reflection, we included the {\tt pcfabs} model 
for representing the partial covering absorption (model C).
The fit gave a similar $\chi^2$ value (the difference of $\chi^2$ is
13) to that for the model B (the 3rd and 4th panels in figure \ref{specabc}).
The best-fit model gave an column density
of $(2.7\pm1.1)\times10^{23}$ cm$^{-2}$
and a covering fraction of 9$\pm$3\% for the partial covering absorber.

We applied the above models to other observations.
Since the observational detector position of the Cen A is different
between 2005 and 2009, a relative normalization between the XIS and PIN
is different.
Unlike 2009, the position of the 2005 observation is not nominal and
thus a nominal relative ratio of normalizations is not available.
Therefore, we at first set 
a relative normalization between the XIS and PIN 
to be free, and obtained it to be 1.06.
This is consistent with the value in Markowitz et al. (2007).
Accordingly, we hereafter fixed the relative normalization to 1.06 for
the 2005 observation.

\begin{figure}[htb]
\centering
\includegraphics[width=0.5\textwidth]{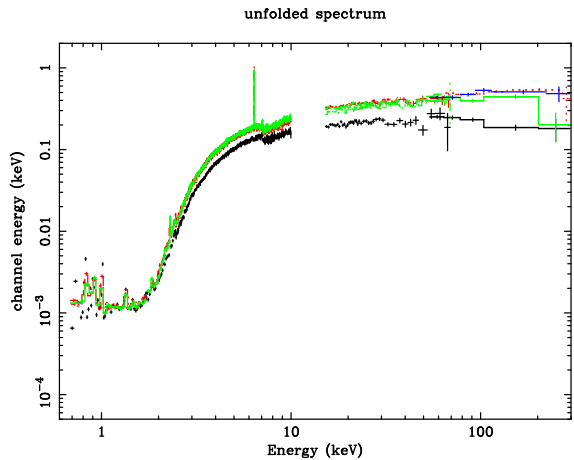}
\caption{$\nu F\nu$ plots of the XIS/PIN/GSO spectra for all
 observations. Black, red, green, blue data correspond to the spectrum
 of the 2005, 2009 1st, 2009 2nd, 2009 3rd observation, respectively.}
\label{speccmpall}
\end{figure}

\begin{figure}[htb]
\centering
\includegraphics[width=0.5\textwidth]{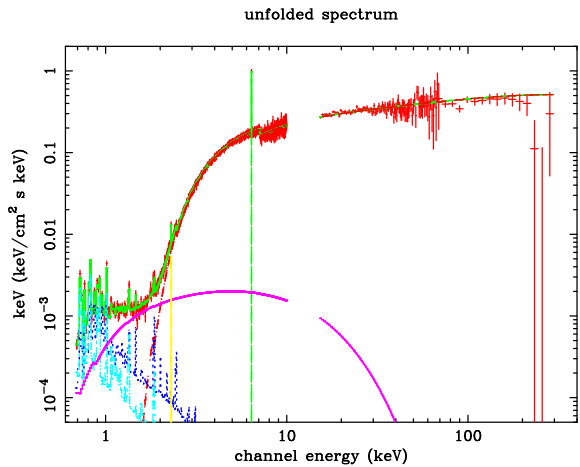}
\caption{$\nu F\nu$ plots of the XIS/PIN/GSO spectra with the model
 C for the 2009 2nd observation.}
\label{specmcsed}
\end{figure}

The fitting results of four observations for the model A, B, and C are
summarized in table \ref{fita}--\ref{fitc}.
The model B and C gave a better fit than the model A for all the
observations, 
and they gave a similar $\chi^2$ value for 2009 three observations, while the
model C gave a significant improvement of the fitting for the 2005
observation.
Markowitz et al. (2007) reported a similar trend for the 2005
observation.
Figure \ref{speccmpall} 
compares the $\nu F\nu$ spectra of all observations, and
Figure \ref{specmcsed} 
shows the best-fit $\nu F\nu$ spectra of the 2009 2nd
observation. 
It can be seen that the soft X-ray emission in 0.7--2 keV is at almost the same
flux level because the emission comes from the extended region, 
while the hard X-ray emission brightened in 2009; 0.2 keV
cm$^{-2}$ s$^{-1}$ and 0.2 keV cm$^{-2}$ s$^{-1}$ at 20 keV and 100 keV,
respectively, 
in 2005, 0.3--0.4 keV
cm$^{-2}$ s$^{-1}$ and 0.4--0.5 keV cm$^{-2}$ s$^{-1}$ at 20 keV and 100 keV,
respectively, in 2009.
In the following, let us look at the fitting results for the model C.
The column density of uniform absorber is almost constant at
$1.0\times10^{23}$ cm$^{-2}$ for all the observations.
The covering fraction of the partial covering absorber is around 30\% in 2005, 
while it is around 8--10\% in 2009.
The column density of the partial covering absorber is not constant, but 
two observations in 2009 gave a Compton-thick value of $>10^{24}$ cm$^{-2}$.
The large difference appears for the powerlaw photon index; 1.94 in 2005 and
1.68--1.71 in 2009.
In other words, the Cen A spectrum became significantly harder in
2009 as can be seen in figure \ref{specmcsed}.
The intensity of the neutral Fe-K fluorescence line is
$(2.3\pm0.2)\times10^{-4}$ c s$^{-1}$ cm$^{-2}$ in 2005 and 
$(2.7-2.9)\times10^{-4}$ c s$^{-1}$ cm$^{-2}$ in 2009.
This is the first evidence of variability of the Fe-K line intensity for
the Cen A, and the Fe-K line intensity is considered to be high 
in 2009 as a result of 
brightening of the Cen A continuum emission since 2007.

Although the time-averaged spectra can be fitted with the model B or C,
both of the models have a problem.
For the model B, the reflection component is generally considered to be 
constant and thus
only the powerlaw component cannot explain the complex time variability 
as described in \S3.1.
The fluorescence Fe-K line is significantly detected with an equivalent
width (EW) 
of 78, 56, 69, 56 eV for the 2005, 2009 1st, 2009 2nd, 2009 3rd
observation, respectively, 
indicating that there is an underlying reflection continuum
and thus the model C does not match this evidence.
The reflection fraction in the model B is around 0.19--0.34 in 2009 and 
0.41 in 2005.
The EW of the Fe-K line against the reflection component is 1.0--2.0
keV, and this is reasonable for the Compton-thick reflector
with one solar abundance.
Therefore, both of the reflection component and the partial covering
absorption are likely to be required.

The above results are obtained by assuming 
that the exponential cut-off energy of the powerlaw model is 1000 keV.
Then, we let it free for the model C, but only a lower limit of
$\sim500$ keV is obtained both in 2005 and 2009.
Therefore, hereafter, we fixed the cut-off energy to 1000 keV.

\subsection{Difference Spectra between High and Low Flux Periods}

Before performing the spectral fitting by including both of the reflection
and the partial covering absorption, it is important to study the
spectral component which produces the complex time variability.
Here, we investigated the difference spectra between high and low flux
periods in 2009 observations.
We used the XIS-F and HXD-PIN data for this analysis, since the XIS-B
and HXD-GSO did not give an enough signal-to-noise ratio for this
analysis.

\begin{figure}[htb]
\centering
\includegraphics[width=0.5\textwidth]{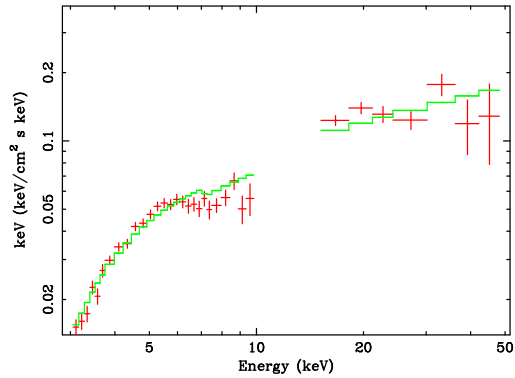}
\caption{$\nu F\nu$ plot of the difference spectra of the XIS/PIN
 between high and low periods in the 2009 1st observation. The solid
 line represents a powerlaw model with a simple absorption.}
\label{hldiff}
\end{figure}

\begin{figure}[htb]
\centering
\includegraphics[width=0.5\textwidth]{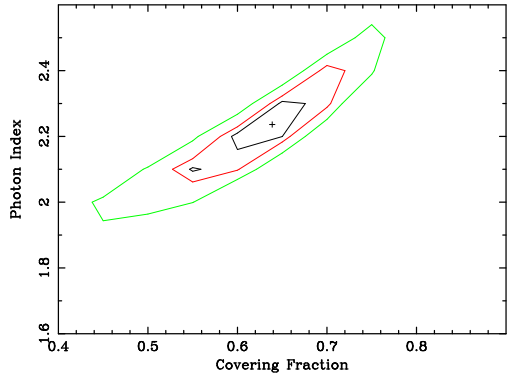}
\caption{Confidence contour between the partial covering fraction of
absorption and powerlaw photon index.}
\label{difspcont}
\end{figure}

First, we took a difference spectrum between high and low flux periods in each
observation.
We defined high and low flux periods in figure \ref{lc} as follows:
8--11th bin, 22--24th bin, and 39--42th bin for high flux
periods, and 1--4th bin, 15--17th bin, and 29--32th for low flux
periods.
We fitted the difference spectra defined as above 
with an absorbed powerlaw model, which 
typically represents the difference spectra of Seyfert galaxies 
(e.g. Miniutti et al. 2007; Shirai et al. 2008).
Figure \ref{hldiff} shows the fitting results for the 1st observation.
Overall spectral shape is represented by the
absorbed powerlaw model, but the fit is not good with a reduced $\chi^2$
value of 2.33. 
There is a strong Fe-K edge feature in the
observed spectrum and it is not reproduced well by the best-fit model.
This indicated a thicker absorber, and thus we tried a partial covering
absorption model.
Then, the fit improved with $\Delta\chi^2=36$, 
and the edge feature can be fitted with this model.
Figure \ref{difspcont} shows a confidence contour between the partial
covering fraction and photon index.
The Compton-thick absorber of $N_{\rm H}=2.6\times10^{24}$ cm$^{-2}$ 
with a high covering fraction of 0.64 is required by the deep Fe-K edge
structure.
The powerlaw photon index is 2.23$\pm$0.18, somewhat steeper than that
for the time-averaged spectra.

For other two observations, since the signal-to-noise ratio of spectra is not
high, the Fe-K edge-feature is less clearer.
However, when the spectra are fitted by a powerlaw model with a
simple absorption, 
the photon index becomes very small; $1.4\pm0.2$ and $1.2\pm0.2$ for
the 2nd and 3rd observation, respectively.
When the partial covering absorption model is introduced, the powerlaw
photon index becomes reasonable around 2 for the 2nd observation.
This is not the case for the 3rd observation; errors of photon index and
absorption column density are large.
When we fixed the photon index to 2.0 for the 3rd observation, 
the absorption column density of the uniform absorber
becomes reasonable around $1\times10^{23}$ cm$^{-2}$.
The fitting results are summarized in table \ref{fitdiff}.
The partial absorber has a large column density of $>10^{24}$ cm$^{-2}$
with a large covering fraction of $>0.3$,
and they seem variable.
These behaviors could cause a complex time variability.
Such a Compton-thick absorber was for the first time observed for the
Cen A.

The partial covering absorber is thought to be variable during each
observation.
Therefore, next, 
we investigated a spectral variability with a finer time resolution.
We divided one observation into several periods with a step of 10 ks, 
following the light curve in figure \ref{lc}, 
and took a difference spectrum between two periods, one of which is the
low flux period just before brightening and the other is around the
highest flux level.
We took the difference between 7th and 3rd bin during the 1st observation, 
10th and 9th during the 1st
observation, 24th and 16th bin during the 2nd observation, 41th and 38th
bin during the 3rd observation in figure \ref{lc}.
Table \ref{fitdiff} 
summarized the fitting results of 
the partial covering absorption model with the photon index fixed to
2.0.
Since the signal-to-noise ratio of each spectrum is not so high, 
errors are large.
However, the column density and covering fraction are not constant while
the column density of the uniform absorber is constant around $1\times10^{23}$ cm$^{-2}$.

In summary, the complex time variability of the Cen A is likely to be 
caused by the variability of the partial covering absorber with a time
scale of $<10000$ sec, while the powerlaw photon index is almost stable
around 2.0.

\subsection{Additional Hard Powerlaw Component}

Since the partial covering absorber is found to exist by the study
of spectral variability, both of the reflection and the partial covering
absorber are needed to fit the Suzaku wide-band X-ray spectra of the
Cen A.
Then, we fitted the time-averaged spectrum of each observation by
considering both of spectral features (model D).
The fitting results are summarized in table \ref{fitd}.
The reduced $\chi^2$ is almost the same as that of the model C (partial
covering absorption without reflection), and the reflection
continuum is not required.
This is inconsistent with the existence of the Fe-K line.
Another problem is that the powerlaw photon index of 1.7 for the
time-averaged spectra in 2009 is different from that obtained
by the analysis of the difference spectra, which require the photon
index of $\sim$2.

The powerlaw component is generally thought to be due to thermal Comptonization
of disk photons, as well as the low/hard state of black 
hole binaries.
For Cyg X-1 and GX 339-4, when they are
brighter in the low/hard state, the Compton thickness becomes larger
but the Comptonizing electron temperature becomes lower (Makishima et
al. 2008, Zdziarski et al. 2004, Del Santo 2008).
As a result, the powerlaw photon index and cut-off energy are larger
and lower, respectively; 
in other words, the spectrum becomes softer in the brighter phase.
A bright Seyfert galaxy, NGC 4151, also exhibited such a trend (Lubi${\rm \acute{n}}$ski
et al. 2010).
On the other hand, the trend of the Cen A is opposite; 
the spectrum became harder in the bright phase.

\begin{figure}[htb]
\centering
\includegraphics[width=0.5\textwidth]{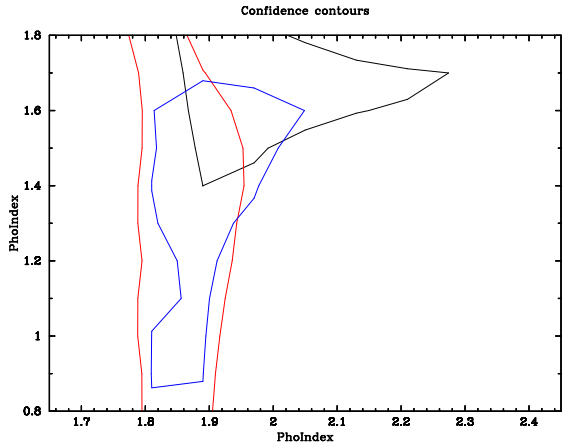}
\caption{Confidence contour between the soft and hard powerlaw photon
 index for fitting with the model E. The contours represent a 95\%
 confidence level for each observation. 
Black solid, red dashed, and blue dotted lines
 correspond to those of 2009 1st, 2nd, and 3rd observation, respectively.}
\label{contpo2}
\end{figure}

\begin{figure}[htb]
\centering
\includegraphics[width=0.5\textwidth]{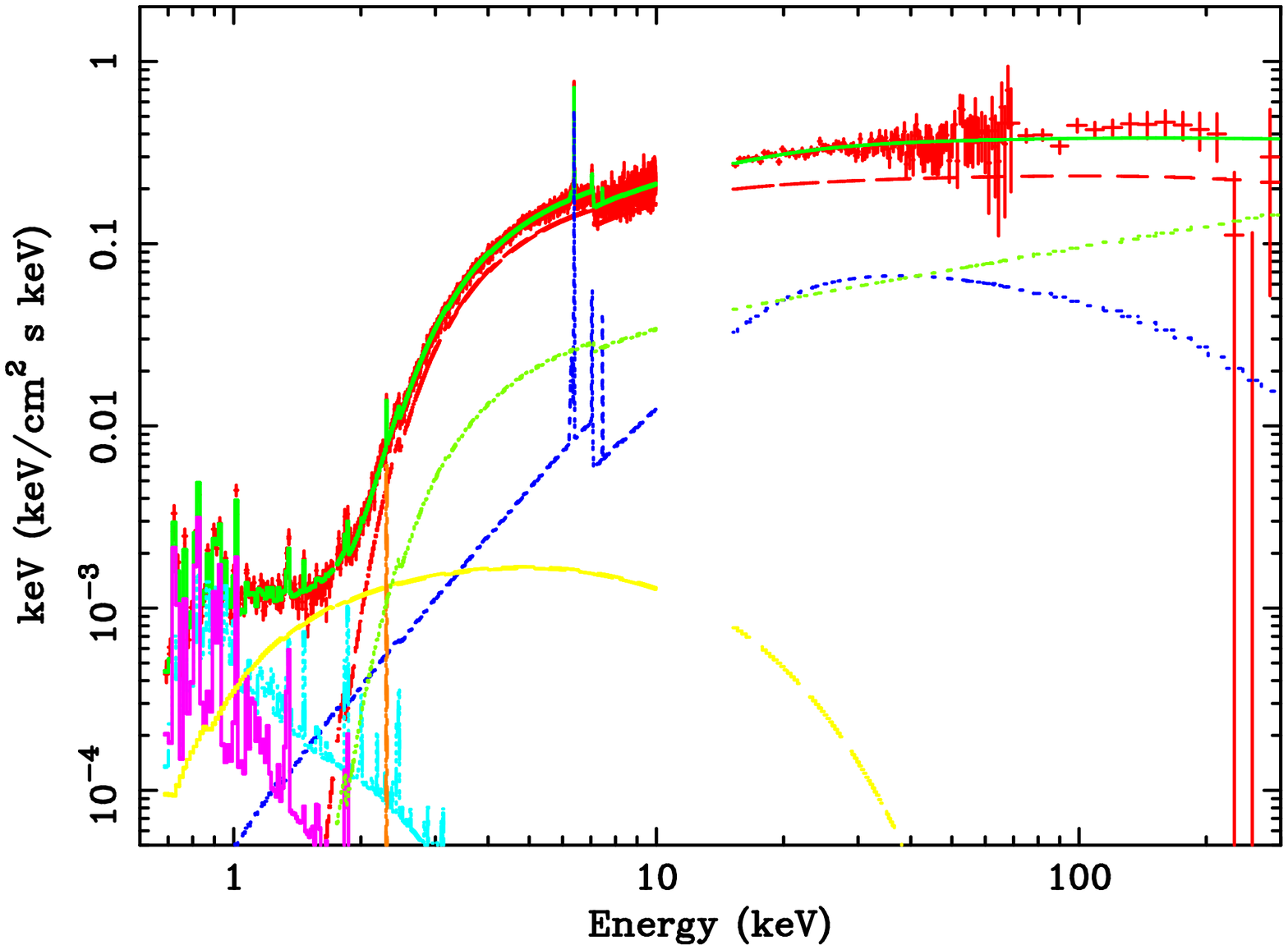}
\caption{$\nu F\nu$ plot of the XIS/PIN/GSO spectra of the 2009 2nd
 observation with the model E.}
\label{specmde}
\end{figure}

The Cen A has been found to be a gamma-ray emitter up to the TeV band,
while black hole binaries and NGC 4151 are not.
Therefore, the nonthermal jet component, which connects to the high
energy gamma-ray band, is expected to exist in the X-ray band.
We included one more powerlaw model in addition to the model D, 
and multiplied it by the absorption with a column density 
of $1\times10^{23}$ cm$^{-2}$;
\textcolor{black}{
the absorption is needed for this additional powerlaw component 
to be below the observed flux in the soft X-ray band, and the above
value is just assumption.
Later we describe about the dependence of results on this absorption.
}
In this case, we cannot constrain the parameters of both powerlaw
components well.
In addition, the intensity of the hard powerlaw component becomes
coupled with that of the reflection component.
Then, we replaced the {\tt pexrav} model by the {\tt pexmon} model 
(Nandra et al. 2007) for the reflection component.
The {\tt pexmon} model considered the Fe-K and Ni-K fluorescence lines
and thus we can constrain the reflection component by the prominent Fe-K
line.
In this case, the gaussian model for the Fe-K line is not included 
for fitting (model E).

Figure \ref{contpo2} shows a confidence contour between the soft and
hard powerlaw photon index, where we plot the 95\% confidence contours
of three 2009 observations.
The photon index of the soft and hard component is constrained to be 
around 1.9 and 1.6, respectively, for all 2009 observations.
The photon index of the hard component depends on the assumed
absorption; when the absorption is weaker, the photon index becomes
smaller and the fraction of the hard component in the softer X-ray band
becomes smaller.
Thus the photon index of 1.6 is considered to give an upper limit of 
the hard component.
Then, we fixed the photon indices of the two powerlaw components to 1.6
and 1.9.
Table \ref{fite} summarized the fitting results, and figure \ref{specmde}
shows the best-fit model and spectra.
The $\chi^2$ value is smaller than those of the model D for the 2009
1st observation, but almost the same as those of the model D for others.
The fraction of the reflection component is around 0.4--0.5 and the Fe
abundance is around 0.7--0.9 solar,
These are typical values for Seyfert galaxies, and therefore reasonable
values; it can be said that we correctly model the reflection component.
The flux of the additional powerlaw component was $<$10\% and 30\% of the
original powerlaw at 100 keV in 2005 and 2009, respectively.
Therefore, the flux increase of the additional hard powerlaw
component, possibly associated with the jet, can explain the harder spectra 
in 2009, but the Seyfert-like
component is still dominant in the Suzaku X-ray band even if the 
jet component exists.


\subsection{Time History of Spectral Parameters}

In order to check the view of the model E, we performed time-resolved
spectral fittings for the 2009 observations.
We divided the XIS, PIN, and GSO data of each observation into 11--14 
periods as a step of 10 ks.
Each period corresponds to each bin of the light curve in figure \ref{lc}.
Since the signal-to-noise ratio is low, 
we fixed the following parameters to the best-fit values obtained for the
time-averaged spectra of each observation in table \ref{fite}; 
reflection fraction, normalization of reflected powerlaw emission, 
Fe abundance of absorber
and reflector, normalization of the S-K line, parameters of the 
soft thermal components.
Then, free parameters are a column density of the uniform absorber, a
column density and covering fraction of the thicker absorber, and
normalizations of two powerlaw component.
In addition, we ignored the data below 3 keV, where the thermal
component is dominant.

\begin{figure}[htb]
\centering
\includegraphics[width=0.5\textwidth]{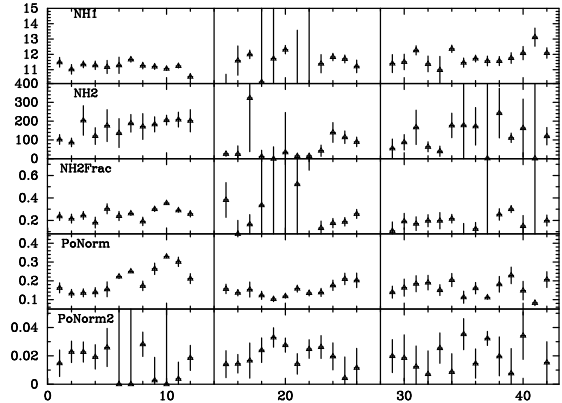}
\caption{Time history of the spectral parameters of the model E for the
 10ks-time-resolved spectra in 
2009 observations. The horizontal number corresponds to the time bin
number in figure \ref{lc}. From top to bottom, an uniform absorption
 column density in unit of $10^{22}$ cm$^{-2}$, a partial covering absorption
 column density in unit of $10^{22}$ cm$^{-2}$, a partial covering
 fraction of absorption, and normalizations of the soft and hard powerlaw
 component in unit of ph cm$^{-2}$ s$^{-2}$ keV$^{-1}$ at 1 keV, are shown.}
\label{specparhist}
\end{figure}

Figure \ref{specparhist} shows a time history of spectral parameters.
The column density of the uniform absorber is less variable; with at most
10\% variability.
The most variable parameters are the normalization of the lower energy
powerlaw component, the column density and covering fraction of the
thicker absorber.
This could create a complex correlation behavior in subsection 3.1.
The normalization of the higher energy powerlaw component is almost
constant, but
a small variability is seen for the 3rd observation. 
It shows an
anticorrelation with the normalization of the lower energy powerlaw
component, and therefore this could be artificial.
No clear correlation between the lower and higher energy powerlaw
components suggests a different origin between two components.

\section{Discussion}

In summary, we measured the broad-band X-ray spectral
variability of the Cen A more accurately than ever with Suzaku, 
and found that the
variable component is a powerlaw with a partially covering
Compton-thick absorption of $\sim10^{24}$ cm$^{-2}$.
We also found a variability of the Fe-K line intensity 
from 2005 to 2009 by a factor of 1.3 or so.
The reflection component associated with the Fe-K line is also
suggested in the spectral modeling, and furthermore an additional hard
powerlaw component with a photon index of $<1.6$ is inferred.

\subsection{Origin of X-ray Time Variability}

X-ray time variability of the Cen A has been reported with the
CGRO/BATSE, CGRO/OSSE, RXTE, XMM-Newton, INTEGRAL, and Swift/BAT, 
for various time scales from sub-days to years (Kinzer et al. 1995, 
Wheaton et al. 1996, 
Rothschild et al. 1999, Rothschild et al. 2006, Evans et al. 2004).

RXTE and INTEGRAL observations reported that the 
time variability is caused by the change of absorption column density, 
the powerlaw photon index, and the powerlaw normalization, based on the
spectral fitting by a powerlaw model with an uniform absorption
(Rothschild et al. 2006).
The absorption column density is in the range of
$(0.9-1.7)\times10^{23}$ cm$^{-2}$, the powerlaw photon index is in the
range of 1.65--1.85, and the flux is $(4-10)\times10^{-10}$ erg cm$^{-2}$ in
20--100 keV.
For the Suzaku observations (table \ref{fita}), 
the column density is $(0.9-1.2)\times10^{23}$ cm$^{-2}$, the powerlaw
photon index is in the range of 1.65--1.82, and 
the powerlaw flux in 20--100 keV is $5.7\times10^{-10}$
erg cm$^{-2}$ s$^{-1}$ and $11\times10^{-10}$ erg cm$^{-2}$ s$^{-1}$ in
2005 and 2009, respectively, based on the same model.
Therefore, the spectral parameters match the past observations.
However, the accurate spectroscopy with Suzaku revealed that a
powerlaw model with an uniform absorption was not valid, 
and the X-ray variation is
partly caused by the change of the partial covering Compton-thick
absorber, together with the change of the powerlaw continuum level with
a time scale of sub-days.
The absorption column density of the uniform one is almost constant
around $1.2\times10^{23}$ cm$^{-2}$.

The variability of the Fe-K line has never been reported in the past 
observations; it is steady
around $5\times10^{-4}$ ph cm$^{-2}$ s$^{-1}$.
Suzaku for the first time confirmed that the Fe-K line intensity
significantly increased by a factor of several tens
percents from 2005 to 2009, following the brightening of the continuum
flux since 2007.
This suggests that the Fe-K line emitter lies at a distance of $<1$ pc from 
the nucleus.
The Fe-K line intensity observed with Suzaku is
$(2.3-3.0)\times10^{-4}$ ph cm$^{-2}$ s$^{-1}$, and therefore it is
relatively weaker than ever.
RXTE results reported that the Fe-K line intensity is around
$5\times10^{-4}$ ph cm$^{-2}$ s$^{-1}$ in 2004, just before the Suzaku
2005 observation.
However, we must take care that the RXTE energy resolution of $\sim$1000
eV cannot accurately resolve a Fe-K line with an EW of $\sim$100 eV.
Chandra and XMM-Newton results
\textcolor{black}{
of $(2-4)\times10^{-4}$ ph cm$^{-2}$ s$^{-1}$ (Evans et al. 2004) 
}
were close to the Suzaku ones.

The behavior in the soft gamma-ray band was well studied with the
OSSE (Kinzer et al. 1995); the spectral cut-off shape varied in such a way that
the cut-off was clearer in the bright phase (0.6 keV cm$^{-2}$ s$^{-1}$ at
100 keV) than in the faint phase (0.2 keV cm$^{-2}$ s$^{-1}$ at 100 keV).
Suzaku 2005 data correspond to the faint phase, and the 2009 data do
to the bright phase.
However, the spectral cut-off is not clearly detected with Suzaku both in 2005
and 2009.
This would be related with the additional powerlaw component, and we discuss
this issue in the following subsection.

\subsection{X-ray Reprocessing Materials}

The location of the stable uniform absorber can be constrained by the
Fe-K edge which is almost attributed to the uniform absorber for the
time-averaged spectra. 
The edge energy is $7.12\pm0.03$ keV in 2009, leading to 
the ionization parameter of 
$\xi=\frac{L}{nR^2}\leq0.1$, where $L$ is the luminosity of the central
engine, $n$ the matter density, and $R$ the distance to the matter
(Kallman et al. 2004).
Considering the column density $N_{\rm H}<nR$,
the uniform absorber lies at $>160$ pc away from the nucleus, where we
take $L=5\times10^{42}$ erg s$^{-1}$ and $N_{\rm H}=1\times10^{23}$
cm$^{-2}$.
Therefore, the uniform absorber is likely to be associated with 
the famous dust lane lying on the elliptical galaxy.

On the other hand, the location of the thicker partial covering 
absorber can be constrained by the Fe-K edge in the difference spectra
between high and low flux states; the edge energy is $<7.3$ keV,
corresponding to $\xi\leq10$.
In the same way as above, using $N_{\rm H}=10^{24}$ cm$^{-2}$, 
the location is constrained to be $R>0.16$ pc.
The variation of absorber parameters with a time scale of 
$\delta t\sim1$ day indicates the blob-like structure.
Taking the black hole mass of $5\times10^7$ $M_{\odot}$, the Kepler
velocity is $v=1.1\times10^8$ cm s$^{-1}$ at a radius of 0.16 pc.
Then, the size of blobs perpendicular to the line of sight becomes 
$v\delta t\sim10^{13}$ cm.
Since the size toward the line of sight is likely to be the same order as
above, the density becomes $n\sim10^{11}$ cm$^{-3}$.

The Fe-K line EW is typical for Seyfert galaxies with a similar
absorption column density (Fukazawa et al. 2011).
This indicates that there is a Compton-thick material covering a large
solid angle of $\sim\pi$ or so.
Considering the variation time scale of $<2$ year, the material is a
molecular torus which is believed to exist commonly in Seyfert galaxies.
The intrinsic emission irradiating the material is unlikely to be the beamed
jet emission, since the jet emission concentrates within a small
solid angle along the jet
direction by the relativistic effect.
Therefore, the Seyfert-like emission originating from
the inner disk region is thought to dominate in the X-ray band.

\subsection{Jet Component}

\begin{figure}[htb]
\centering
\includegraphics[width=0.5\textwidth]{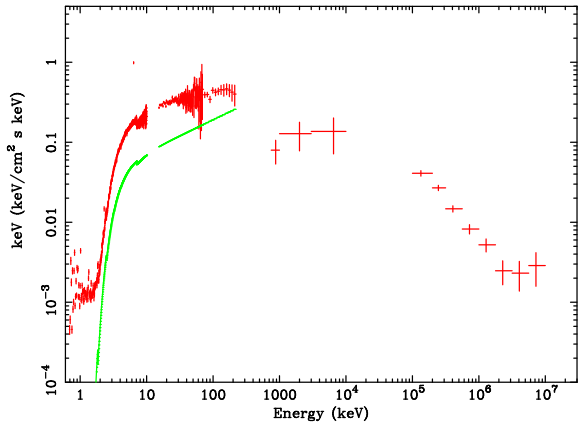}
\caption{$\nu F\nu$ plot of the Cen A from the X-ray to GeV
 gamma-ray band. Data points are obtained with the Suzaku (2009), CGRO/COMPTEL
 (1991--1994), and Fermi/LAT (2008--2009). The solid line represents the
 hard powerlaw component suggested with the Suzaku observation.}
\label{cenased}
\end{figure}

Our analysis of the Suzaku data suggests the additional hard powerlaw
component with a photon index of $<1.6$, whose flux is around
30\% of the total flux at 100 keV.
This hard component might be brighter in 2009 active phase than in 2005
faint phase.
The crossover energy against another powerlaw emission, which is
likely to be a Seyfert-like nuclear emission dominated below 100 keV, 
is around 400 keV.
This hard component seems to smoothly connect to the CGRO/COMPTEL MeV gamma-ray
emission as shown in figure \ref{cenased}.
CGRO/EGRET and Fermi/LAT detected the GeV gamma-ray emission from the
Cen A, whose spectrum well connects to the MeV emission.
As reported by Abdo et al. (2010c), the multi-wavelength spectrum of the
Cen A can be modeled by the Synchrotron Self-Compton model.
The predicted jet flux in the X-ray band 
strongly depends on the model parameters, such as magnetic field,
emission size, low energy electron spectrum, and so on (Abdo et al. 2010c).
Both of decelerating jet (Georganopoulos \& Kazanas 2003) and structure jet
(Chiaberge et al. 2000) can explain the X-ray emission inferred in this
paper, and
the possible hard component obtained by the Suzaku data could be a lower
energy part of the Compton component.

\textcolor{black}{
Evans et al. (2004) suggested the jet emission in the soft X-ray band
with a photon index of around 2 and a flux of 
$\sim1\times10^{-2}$ keV cm$^{-2}$ s$^{-1}$.
This flux is somewhat lower than the hard component suggested with
Suzaku, but could be the same component.
However, as Evans et al. (2004) described, their soft component can be
explained by the leaked nuclear component due to the partial covering
absorber.
Or, since Evans et al. (2004) did not include the reflection continuum
in the spectral model, their soft component could be a reflection
continuum which we considered in the spectral fitting.
Alternatively, their soft component might be synchrotron emission from
jets and the X-ray band covers the transition region from synchrotron
emission to inverse Compton scattering.
Anyway, we cannot conclude whether their origins are the same or not.
}.

Jet emission has been detected in the X-ray band for a radio galaxy 
3C120 with Suzaku 
(Kataoka et al. 2007); where the variable soft X-ray component was detected
with a shorter time scale than the Seyfert-like nuclear emission.
For the Cen A, the possible jet X-ray component has a longer time scale of
variability than the Seyfert-like emission.
Considering that GeV gamma-ray emission shows no significant variability
over one year (Abdo et al. 2010c), the jet emission is less beamed and
thus the relativistic effect on the variability is smaller.
The jet emission in the X-ray band is often not
well understood for other radio galaxies.
ASTRO-H/SGD will give us the first opportunity to detect the hard excess
component clearly from the Cen A and other radio galaxies.
Also, the ASTRO-H can search for the jet emission from the X-ray
spectral variability with the sensitive wide-band X-ray spectroscopy, 
as well as the Cen A and 3C120.

Authors thanks to Dr. J. Kataoka and the anonymous referee for careful
reading and helpful comments. 
The authors also wish to thank all members of the Suzaku Science Working
Group, for their contributions to the instrument preparation, spacecraft
operation, software development, and in-orbit calibration. This work is
partly supported by Grants-in-Aid for Scientific Research by the
Ministry of Education, Culture, Sports, Science and Technology of Japan 
(20340044).

\appendix
\section{GSO background reproducibility for the Cen A data}

\begin{figure}[htb]
\centering
\includegraphics[width=0.5\textwidth]{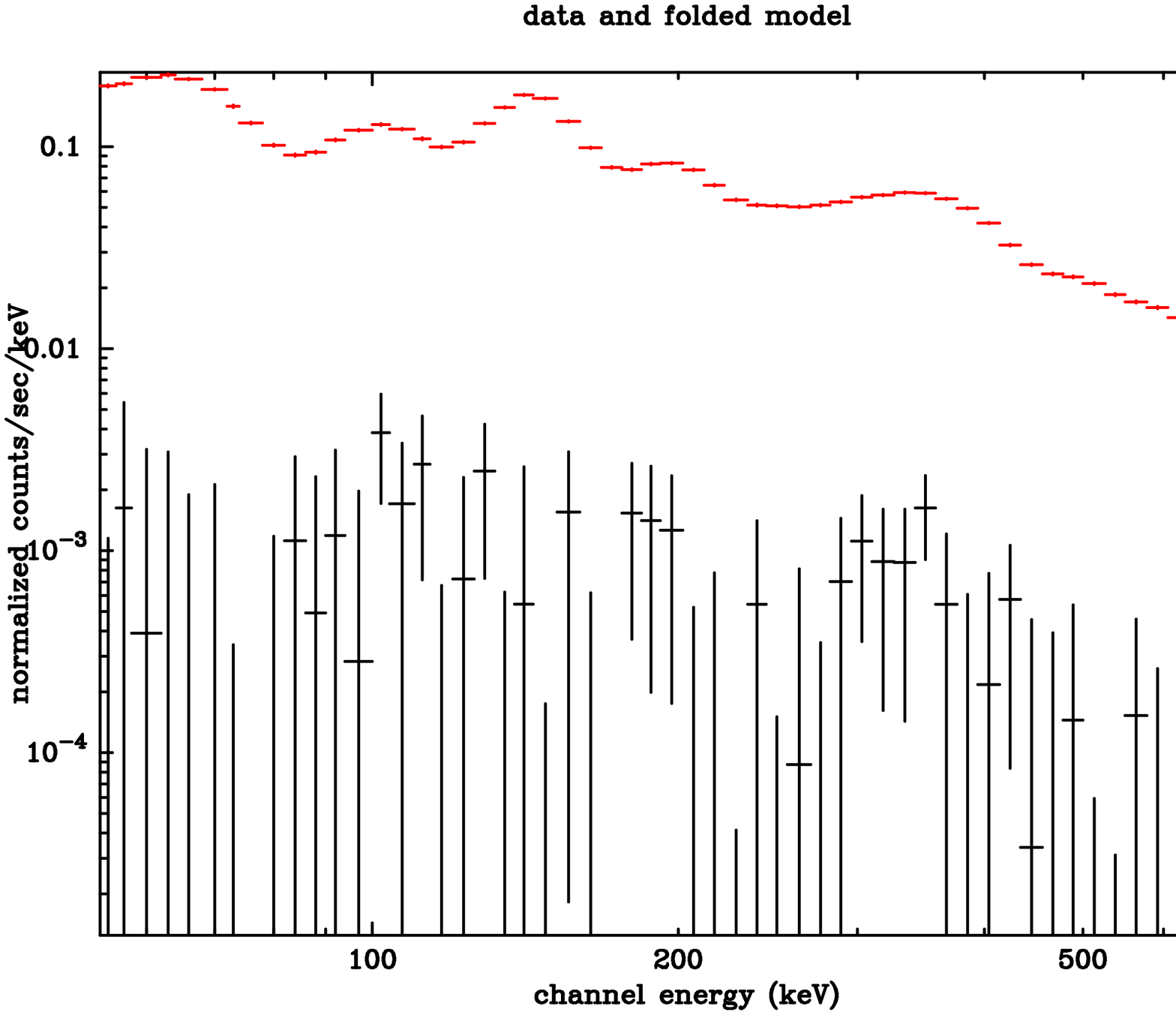}
\includegraphics[width=0.5\textwidth]{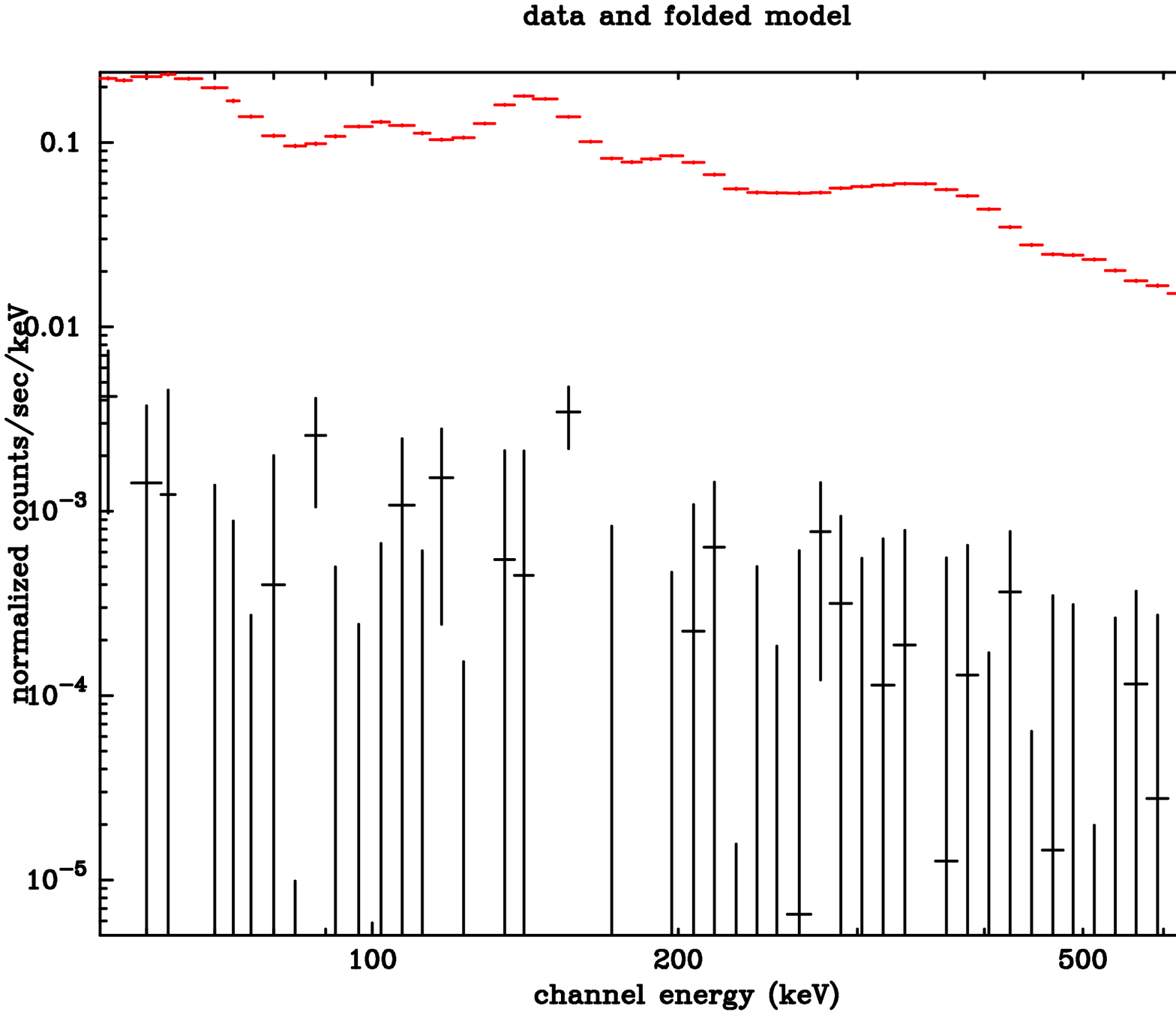}
\includegraphics[width=0.5\textwidth]{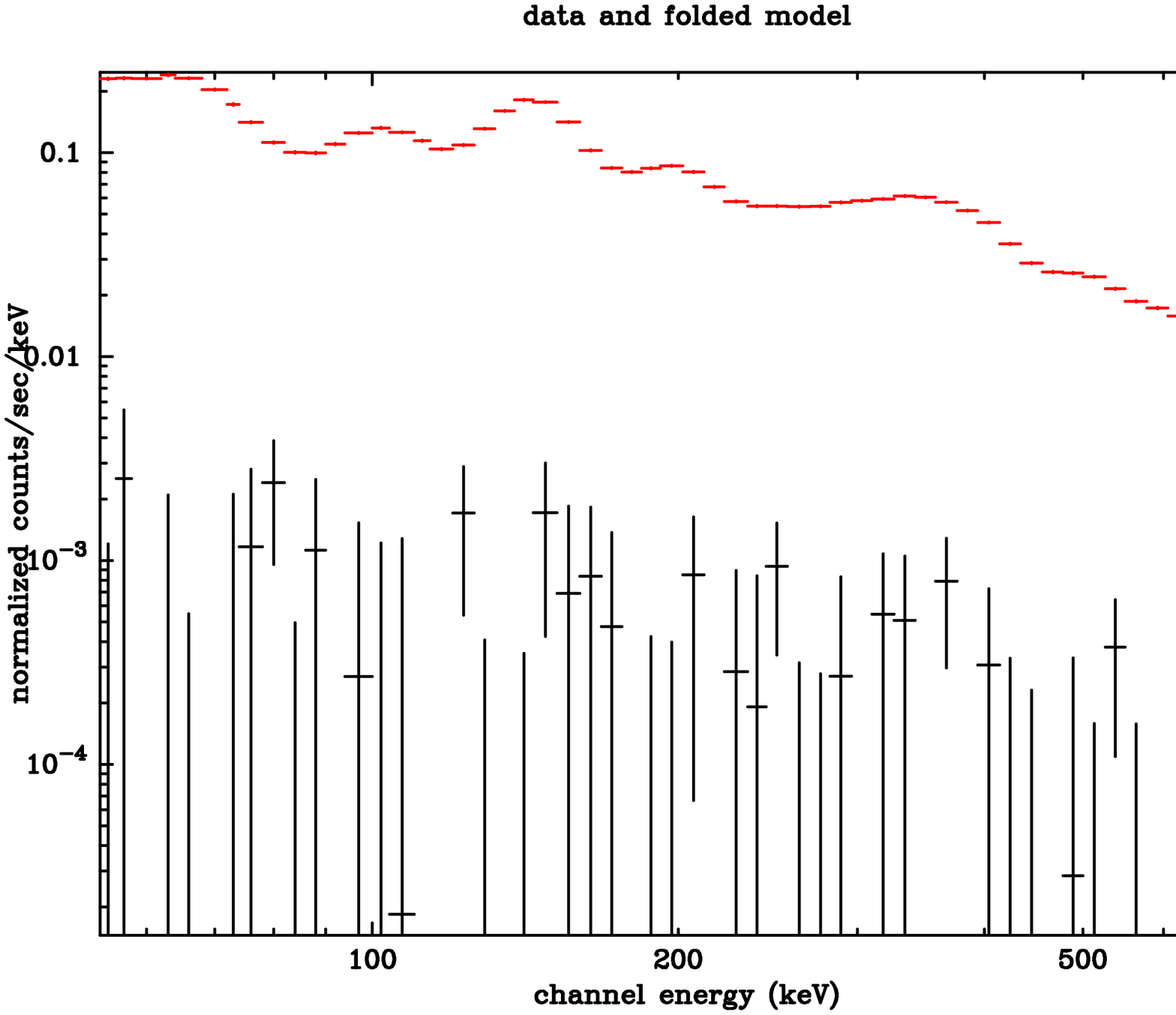}
\caption{Comparison of GSO spectra between the data and background model during the earth
occultation period in each observation.}
\label{earthsp}
\end{figure}

\begin{figure}[htb]
\centering
\includegraphics[width=0.5\textwidth]{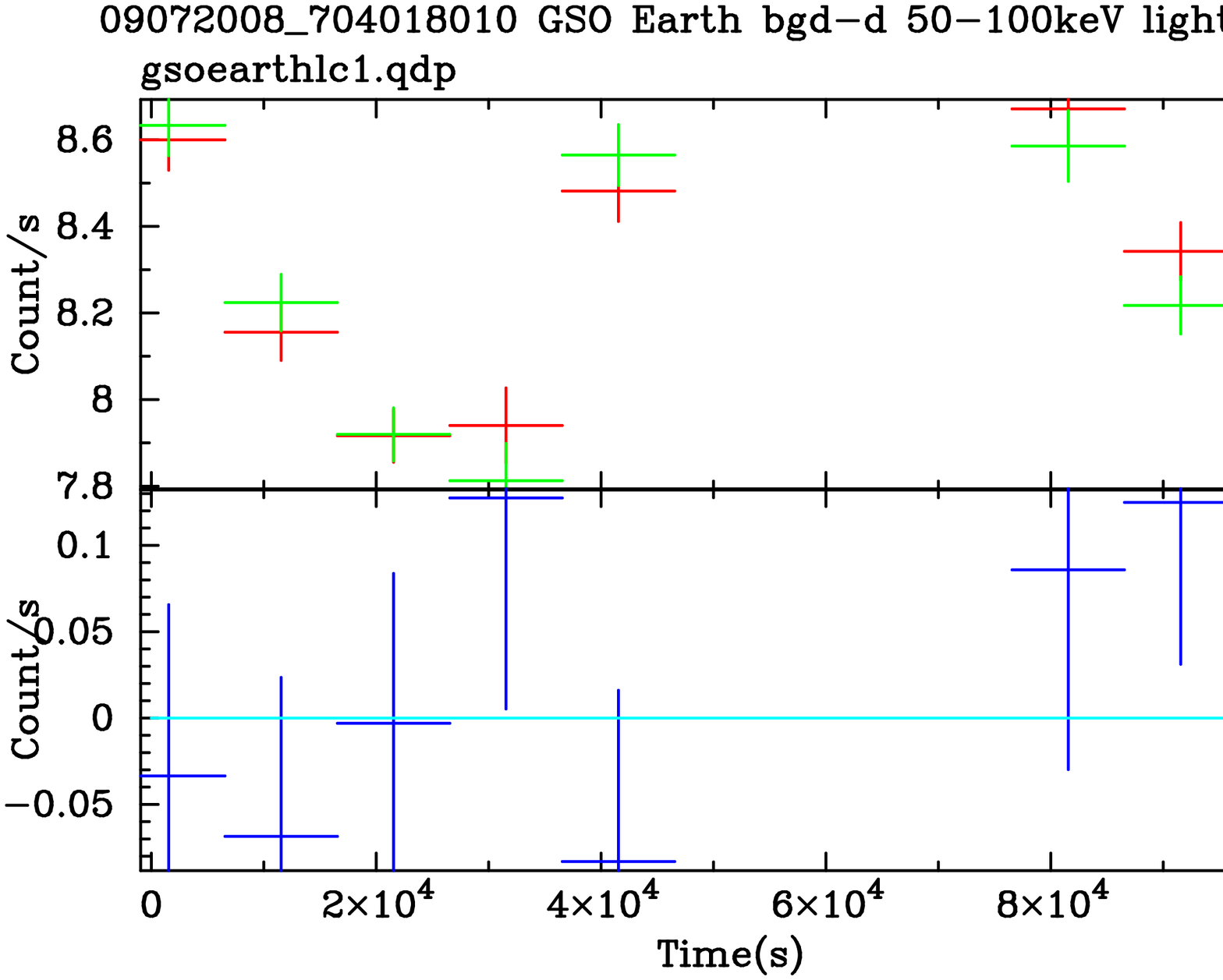}
\includegraphics[width=0.5\textwidth]{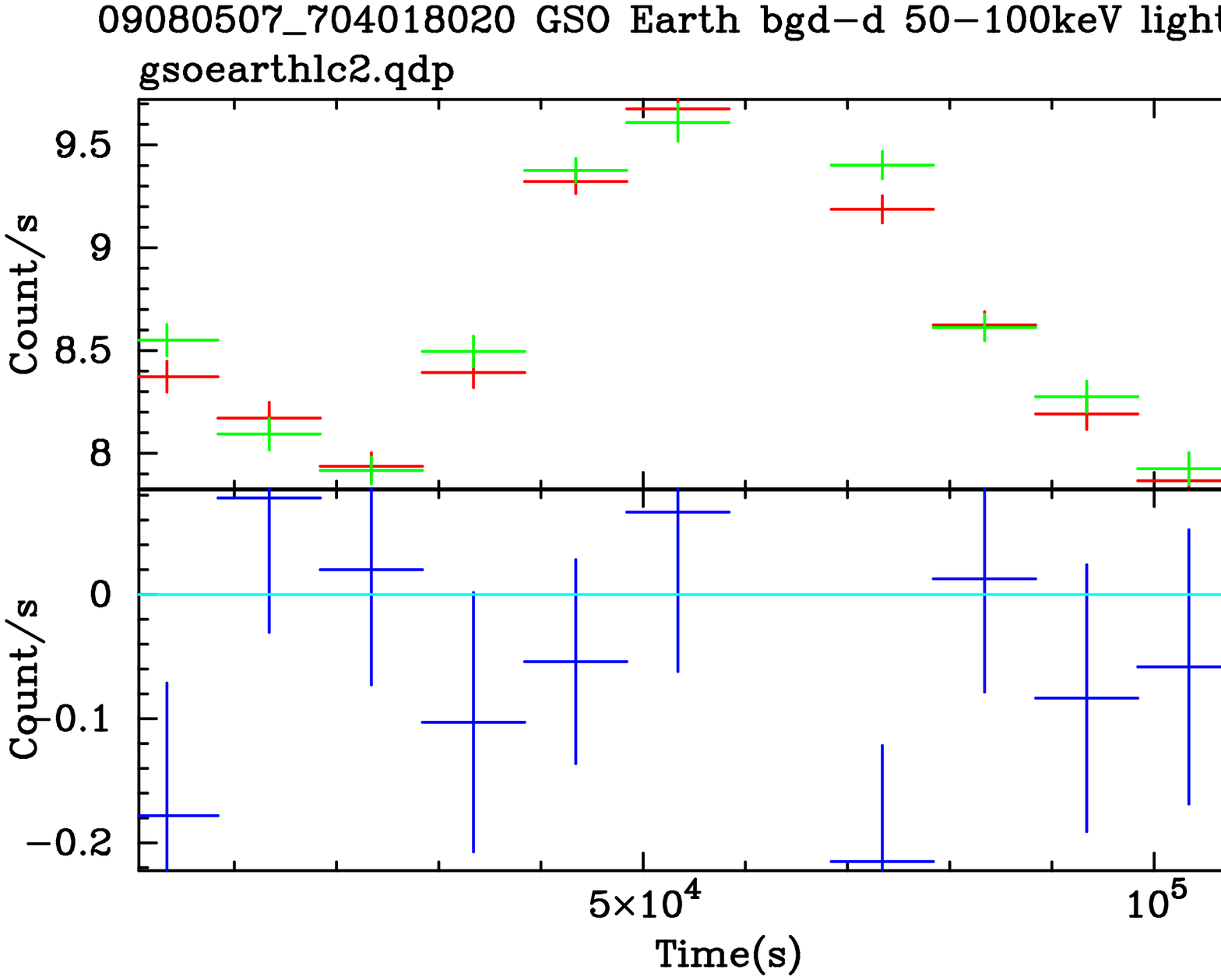}
\includegraphics[width=0.5\textwidth]{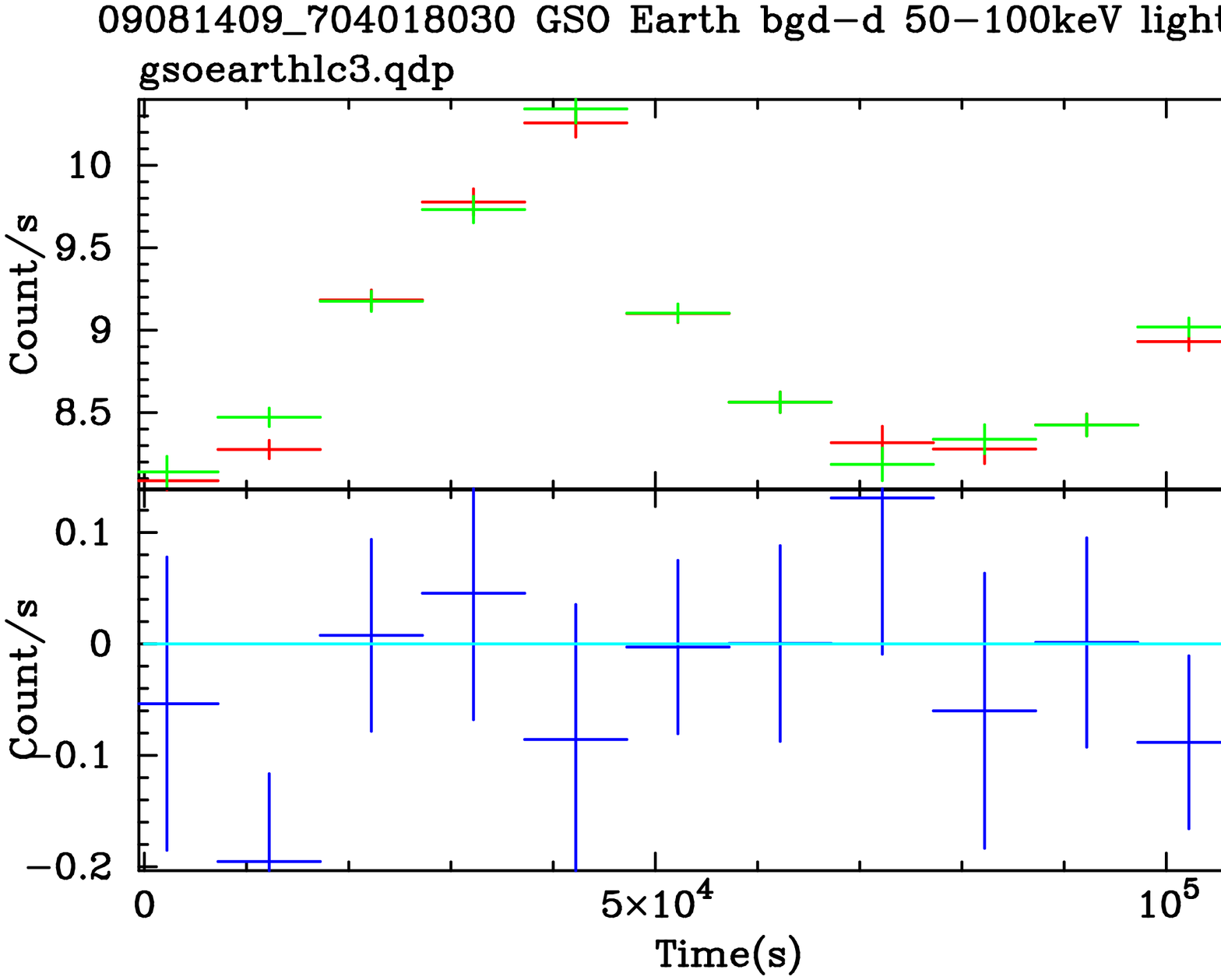}
\caption{Comparison of GSO light curves in 50--100 keV 
between the data and background model during the earth 
occultation period in each observation.}
\label{earthlc}
\end{figure}

Since the GSO signal rate from the Cen A is less than 10\% of
the GSO background, the systematic uncertainty of the GSO background
reproducibility is not ignored.
Then, we checked the reproducibility by using the earth occultation data
during the 2009 Cen A observation.
Since the CXB is negligible, we expect no extra signal than the
background.
Figure \ref{earthsp} shows the comparison of GSO spectra between the
earth occultation data and background model, indicating that the
background model well reproduce the earth occultation spectra with an
accuracy of 2\% or so.
Figure \ref{earthlc} shows the comparison of GSO light curve between the
earth occultation data and background model, indicating that the
background model well reproduce the history of the earth occultation
rate with an accuracy of 2\% or so.
Therefore, the background reproducibility is as good as 1\%.

For the 2005 observation, no earth occultation occurred during the observation.
However, the background rate is lower by a factor of
$\sim2$ than the 2009 observation.
Therefore, the effect of the background uncertainty is expected to be smaller.

\begin{table}[htb]
\caption{Summary of Suzaku observations of the Cen A.}
\label{obs}
\begin{center}
\small
\begin{tabular}{clccc}
\hline\hline
Observation ID & Date & Start & Stop & Exposure$^{\star}$ \\
\hline
100005010 & Aug. 19--20, 2005 & 08-19 03:39:19 & 08-20 09:50:08 & 37984 \\
704018010 & Jul. 20--21, 2009 & 07-20 08:55:29 & 07-21 18:26:24 & 44587 \\
704018020 & Aug. 5--6, 2009 & 08-05 07:23:27 & 08-06 16:52:14 & 35093 \\
704018030 & Aug. 14--16, 2009 & 08-14 09:06:56 & 08-16 02:31:24 & 37636 \\
\hline
\multicolumn{5}{l}{$\star$: Exposure time of th HXD-PIN data after data reduction.}\\
\end{tabular}
\end{center}
\normalsize
\end{table}

\begin{table}[htb]
\caption{Fitting results of the XIS soft thermal components.}
\label{xisapec}
\begin{center}
\small
\begin{tabular}{ccccccc}
\hline\hline
Obs & $I_{\rm bremss}^{a}$ & $kT_1$ & $I_{apec1}^{b}$ & $kT_2$ & $I_{apec2}^{c}$ & $\chi^2/dof$ \\
 & ($10^{-3}$) & (keV) & ($10^{-3}$) & (keV) & ($10^{-3}$) & ($\chi^2,dof$) \\
\hline
2005 & 1.5$\pm$0.1 & 0.71$\pm$0.02 & 1.7$\pm$0.3 & 0.32$\pm$0.01  & 3.9$\pm$0.8 &  2.02 (616.8/306) \\ 
2009a & 1.6$\pm$0.1 & 0.72$\pm$0.03 & 1.8$\pm$0.4 & 0.33$\pm$0.03  & 3.8$\pm$1.2 &  1.49 (455.2/306) \\ 
2009b & 1.8$\pm$0.1 & 0.79$\pm$0.05 & 1.6$\pm$0.5 & 0.29$\pm$0.01  & 5.3$\pm$0.5 &  1.25 (389.9/313) \\ 
2009c & 1.7$\pm$0.1 & 0.72$\pm$0.04 & 2.0$\pm$0.5 & 0.28$\pm$0.02  & 4.9$\pm$1.5 &  1.30 (398.2/306) \\ 
Adapted Values$^d$ & -- & 0.72 & -- & 0.30  & 4.0 & --  \\ 
\hline
\multicolumn{7}{p{15cm}}{Two apec models are multiplied by the photoelectric
 absorption, whose absorption column density is fixed to
 $1.6\times10^{21}$ cm$^{-2}$.}\\
\multicolumn{7}{p{15cm}}{Metal abundances of both apec models are fixed to 0.3 solar. A
 temperature of the bremmstrahlung is fixed to 7 keV.}\\
\multicolumn{7}{l}{$a$: Normalization of the bremmstrahlung.}\\
\multicolumn{7}{l}{$b$: Normalization of the apec model 1.}\\
\multicolumn{7}{l}{$c$: Normalization of the apec model 2.}\\
\multicolumn{7}{l}{$d$: Adapted values of spectral parameters for
 fitting the wide-band Suzaku spectra.}
\end{tabular}
\end{center}
\normalsize
\end{table}

\begin{table}[htb]
\caption{Results of simultaneous fitting of the XIS/PIN/GSO spectra (model A$^{\dagger}$).}
\label{fita}
\begin{center}
\small
\begin{tabular}{cccccc}
\hline\hline
Parameters & & 2005 & 2009a & 2009b & 2009c \\
\hline
$N_{\rm H}^a$ & ($10^{23}$) &	1.20$\pm$0.01 	&	0.93$\pm$0.01 	&	1.03$\pm$0.01 	&	1.04$\pm$0.01 	\\
$A_{\rm Fe}^b$ & (solar) &	0.65$\pm$0.05 	&	1.53$\pm$0.06 	&	1.10$\pm$0.07 	&	1.09$\pm$0.06 	\\
$\alpha_{\rm ph}^c$ & &	1.82$\pm$0.05 	&	1.64$\pm$0.05 	&	1.68$\pm$0.06 	&	1.67$\pm$0.05 	\\
$I_{\rm pow}^d$ & &	0.117$\pm$0.003 	&	0.117$\pm$0.003 	&	0.117$\pm$0.003 	&	0.126$\pm$0.003 	\\
$E_{\rm Fe}^e$ & (keV) &	6.398$\pm$0.004 	&	6.392$\pm$0.006 	&	6.394$\pm$0.007 	&	6.395$\pm$0.008 	\\
$I_{\rm Fe}^f$ & ($10^{-4}$) &	2.6$\pm$0.2 	&	2.6$\pm$0.3 	&	3.0$\pm$0.3 	&	2.7$\pm$0.3 	\\
$\chi^2/dof$ & &	 1.85 	&	 1.50 	&	 1.14 	&	 1.31 	\\
($\chi^2,dof$) & &	(2314.7/1248) 	&	(2093.4/1392) 	&	(1539.8/1353) 	&	(1796.7/1369) 	\\
\hline
\multicolumn{6}{p{15cm}}{$\dagger$: Model A is {\tt constant*phabs*( [highecut*zvphabs*powerlaw + zgauss] + apec + apec + phabs*bremss + zgauss)} in xspec. Components between parentheses [ ] represent the AGN emission.}\\
\multicolumn{6}{p{15cm}}{$a,b$: Hydrogen column density and Fe abundance of
 the absorber. The column density is in unit of cm$^{-2}$ for the {\tt zvphabs} model.}\\
\multicolumn{6}{p{15cm}}{$c,d$: Photon index and normalization at 1 keV of the power-law model.}\\
\multicolumn{6}{p{15cm}}{$e,f$: Center energy and normalization of the
 Fe-K$\alpha$ line. The normalization is in unit of ph cm$^{-2}$ s$^{-1}$.}\\
\end{tabular}
\end{center}
\normalsize
\end{table}

\begin{table}[htb]
\caption{Results of simultaneous fitting of the XIS/PIN/GSO spectra (model B$^{\dagger}$).}
\label{fitb}
\begin{center}
\small
\begin{tabular}{cccccc}
\hline\hline
Parameters & & 2005 & 2009a & 2009b & 2009c \\
\hline
$N_{\rm H}^a$ & ($10^{23}$) &	1.25$\pm$0.01 	&	1.01$\pm$0.01 	&	1.07$\pm$0.01 	&	1.08$\pm$0.01 	\\
$A_{\rm Fe}^b$ & (solar) &	0.46$\pm$0.03 	&	1.14$\pm$0.05 	&	0.94$\pm$0.06 	&	0.86$\pm$0.07 	\\
$\alpha_{\rm ph}^c$ & &	1.90$\pm$0.05 	&	1.73$\pm$0.05 	&	1.73$\pm$0.06 	&	1.72$\pm$0.06 	\\
$I_{\rm pow}^d$ & &	0.127$\pm$0.003 	&	0.132$\pm$0.003 	&	0.126$\pm$0.004 	&	0.131$\pm$0.004 	\\
$E_{\rm Fe}^e$ & (keV) &	6.398$\pm$0.004 	&	6.391$\pm$0.006 	&	6.396$\pm$0.006 	&	6.396$\pm$0.007 	\\
$I_{\rm Fe}^f$ & ($10^{-4}$) &	2.7$\pm$0.2 	&	2.7$\pm$0.3 	&	3.1$\pm$0.3 	&	2.8$\pm$0.3 	\\
$R^g$ & ($\Omega/2\pi$) &	0.406$\pm$0.002 	&	0.345$\pm$0.003 	&	0.192$\pm$0.004 	&	0.220$\pm$0.005 	\\
$\chi^2/dof$ & &	 1.71 	&	 1.42 	&	 1.11 	&	 1.23 	\\
($\chi^2,dof$) & &	(2129.7/1246) 	&	(1968.6/1390) 	&	(1505.0/1351) 	&	(1683.6/1367) 	\\
\hline
\multicolumn{6}{p{15cm}}{$\dagger$: Model B is {\tt constant*phabs*( [highecut*zvphabs*powerlaw + zgauss + phabs*pexrav] + apec + apec + phabs*bremss + zgauss)} in xspec. Components between parentheses [ ] represent the AGN emission.}\\
\multicolumn{6}{p{15cm}}{$a,b$: Hydrogen column density and Fe abundance of 
 the absorber. The column density is in unit of cm$^{-2}$ for the {\tt zvphabs} model.}\\
\multicolumn{6}{p{15cm}}{$c,d$: Photon index and normalization at 1 keV of the power-law model.}\\
\multicolumn{6}{p{15cm}}{$e,f$: Center energy and normalization of the
 Fe-K$\alpha$ line. The normalization is in unit of ph cm$^{-2}$ s$^{-1}$.}\\
\multicolumn{6}{p{15cm}}{$g$: Fraction of the reflection component for the  {\tt pexrav} model.}\\
\end{tabular}
\end{center}
\normalsize
\end{table}

\begin{table}[htb]
\caption{Results of simultaneous fitting of the XIS/PIN/GSO spectra (model C$^{\dagger}$).}
\label{fitc}
\begin{center}
\small
\begin{tabular}{cccccc}
\hline\hline
Parameters & & 2005 & 2009a & 2009b & 2009c \\
\hline
$N_{\rm H}^a$ & ($10^{23}$) &	1.17$\pm$0.04 	&	0.97$\pm$0.03 	&	1.01$\pm$0.03 	&	1.05$\pm$0.04 	\\
$A_{\rm Fe}^b$ & (solar) &	0.51$\pm$0.10 	&	1.33$\pm$0.11 	&	1.11$\pm$0.08 	&	0.93$\pm$0.11 	\\
$\alpha_{\rm ph}^c$ & &	1.94$\pm$0.46 	&	1.68$\pm$0.30 	&	1.71$\pm$0.31 	&	1.69$\pm$0.43 	\\
$I_{\rm pow}^d$ & &	0.176$\pm$0.040 	&	0.138$\pm$0.021 	&	0.129$\pm$0.020 	&	0.138$\pm$0.029 	\\
$E_{\rm Fe}^e$ & (keV) &	6.398$\pm$0.005 	&	6.392$\pm$0.006 	&	6.395$\pm$0.007 	&	6.396$\pm$0.007 	\\
$I_{\rm Fe}^f$ & ($10^{-4}$) &	2.3$\pm$0.2 	&	2.7$\pm$0.3 	&	2.9$\pm$0.3 	&	2.8$\pm$0.3 	\\
$N_{\rm H,2}^g$ & ($10^{24}$) &	0.41$\pm$0.04 	&	1.58$\pm$0.33 	&	0.27$\pm$0.11 	&	1.10$\pm$0.27 	\\
$f_{\rm NH2}^h$ & &	0.29$\pm$0.03 	&	0.10$\pm$0.02 	&	0.09$\pm$0.03 	&	0.08$\pm$0.03 	\\
$\chi^2/dof$ & &	 1.43 	&	 1.42 	&	 1.11 	&	 1.23 	\\
($\chi^2,dof$) & &	(1783.8/1245) 	&	(1970.6/1390) 	&	(1494.2/1351) 	&	(1677.6/1367) 	\\
\hline
\multicolumn{6}{p{15cm}}{$\dagger$: Model C is {\tt constant*phabs*( [pcfabs*highecut*zvphabs*powerlaw + zgauss] + apec + apec + phabs*bremss + zgauss)} in xspec. Components between parentheses [ ] represent the AGN emission.}\\
\multicolumn{6}{p{15cm}}{$a,b$: Hydrogen column density and Fe abundance of 
 the absorber. The column density is in unit of cm$^{-2}$ for the {\tt zvphabs} model.}\\
\multicolumn{6}{p{15cm}}{$c,d$: Photon index and normalization at 1 keV of the power-law model.}\\
\multicolumn{6}{p{15cm}}{$e,f$: Center energy and normalization of the
 Fe-K$\alpha$ line. The normalization is in unit of ph cm$^{-2}$ s$^{-1}$.}\\
\multicolumn{6}{p{15cm}}{$g,h$: Hydrogen column density and covering raction of the absorber for the {\tt pcfabs} model.}\\
\end{tabular}
\end{center}
\normalsize
\end{table}

\begin{table}[htb]
\caption{Results of spectral fitting of the XIS/PIN difference spectra
 with a powerlaw model with a partial covering absorption.}
\label{fitdiff}
\begin{center}
\small
\begin{tabular}{lccccc}
\hline\hline
Observation & & $N_{\rm H1}^a$ & $\alpha^b$ & $N_{\rm H2}^c$ & $f_{\rm NH2}^d$ \\
 & & $10^{22}$ & & $10^{24}$ & \\
\hline
2009 1st & High--Low & $13.2\pm1.3$ & $2.23\pm0.18$ & $2.67\pm0.98$ & $0.64\pm0.09$ \\
2009 2nd & High--Low & $11.7\pm6.5$ & $2.08\pm0.81$ & $>1.00$ & $0.65\pm0.43$ \\
2009 3rd & High--Low & $11.6\pm0.7$ & 2 (fix) & $>0.51$ & $0.72\pm0.17$ \\
\hline
2009 1st & 7--3th & $13.6\pm2.1$ & 2 (fix) & $>0.57$ & $0.29\pm0.21$ \\
2009 1st & 10--9th & 12(fix) & 2 (fix) & $>1.19$ & $0.64\pm0.15$ \\
2009 2nd & 24--16th & $11.4\pm3.0$ & 2 (fix) & $>1.31$ & $0.71\pm0.13$ \\
2009 3rd & 41--38th & 12(fix) & 2 (fix) & $>0.03$ & $0.60\pm0.25$ \\
\hline
\multicolumn{6}{p{15cm}}{$a$: Hydrogen column density of the uniform
 absorber in unit of cm$^{-2}$.}\\
\multicolumn{6}{p{15cm}}{$b$: Photon index and normalization at 1 keV of the power-law model.}\\
\multicolumn{6}{p{15cm}}{$c,d$: Hydrogen column density and covering raction of the absorber for the {\tt pcfabs} model.}\\
\end{tabular}
\end{center}
\normalsize
\end{table}

\begin{table}[htb]
\caption{Results of simultaneous fitting of the XIS/PIN/GSO spectra (model D$^{\dagger}$).}
\label{fitd}
\begin{center}
\small
\begin{tabular}{cccccc}
\hline\hline
Parameters & & 2005 & 2009a & 2009b & 2009c \\
\hline
$N_{\rm H}^a$ & ($10^{23}$) &	1.17$\pm$0.03 	&	1.01$\pm$0.01 	&	1.07$\pm$0.02 	&	1.08$\pm$0.01 	\\
$A_{\rm Fe}^b$ & (solar) &	0.50$\pm$0.10 	&	1.13$\pm$0.05 	&	0.92$\pm$0.07 	&	0.83$\pm$0.07 	\\
$\alpha_{\rm ph}^c$ & &	1.96$\pm$0.42 	&	1.73$\pm$0.05 	&	1.73$\pm$0.12 	&	1.71$\pm$0.07 	\\
$I_{\rm pow}^d$ & &	0.169$\pm$0.035 	&	0.134$\pm$0.003 	&	0.128$\pm$0.007 	&	0.133$\pm$0.004 	\\
$E_{\rm Fe}^e$ & (keV) &	6.397$\pm$0.004 	&	6.391$\pm$0.006 	&	6.396$\pm$0.007 	&	6.396$\pm$0.007 	\\
$I_{\rm Fe}^f$ & ($10^{-4}$) &	2.4$\pm$0.2 	&	2.7$\pm$0.3 	&	3.0$\pm$0.3 	&	2.8$\pm$0.3 	\\
$N_{\rm H,2}^g$ & ($10^{24}$) &	0.33$\pm$0.04 	&	1.64$\pm$1.08 	&	0.34$\pm$0.29 	&	0.98$\pm$0.68 	\\
$f_{\rm NH2}^g$ & &	0.26$\pm$0.03 	&	$<0.05$ 	&	$<0.05$ 	&	$<0.03$ 	\\
$R^i$ & ($\Omega/2\pi$) &	0.236$\pm$0.002 	&	0.304$\pm$0.003 	&	0.147$\pm$0.005 	&	0.183$\pm$0.004 	\\
$\chi^2/dof$ & &	 1.43 	&	 1.41 	&	 1.11 	&	 1.23 	\\
($\chi^2,dof$) & &	(1773.7/1243) 	&	(1962.7/1388) 	&	(1495.2/1349) 	&	(1678.7/1365) 	\\
\hline
\multicolumn{6}{p{15cm}}{$\dagger$: Model D is {\tt constant*phabs*( [pcfabs*highecut*zvphabs*powerlaw + zgauss + phabs*pexrav] + apec + apec + phabs*bremss + zgauss)} in xspec. Components between parentheses [ ] represent the AGN emission.}\\
\multicolumn{6}{p{15cm}}{$a,b$: Hydrogen column density and Fe abundance of 
 the absorber. The column density is in unit of cm$^{-2}$ for the {\tt zvphabs} model.}\\
\multicolumn{6}{p{15cm}}{$a,b$: Hydrogen column density and Fe abundance of 
 the absorber. The column density is in unit of cm$^{-2}$ for the {\tt zvphabs} model.}\\
\multicolumn{6}{p{15cm}}{$c,d$: Photon index and normalization at 1 keV of the power-law model.}\\
\multicolumn{6}{p{15cm}}{$e,f$: Center energy and normalization of the
 Fe-K$\alpha$ line. The normalization is in unit of ph cm$^{-2}$ s$^{-1}$.}\\
\multicolumn{6}{p{15cm}}{$g,h$: Hydrogen column density and covering raction of the absorber for the {\tt pcfabs} model.}\\
\multicolumn{6}{p{15cm}}{$i$: Fraction of the reflection component for the  {\tt pexrav} model.}\\
\end{tabular}
\end{center}
\normalsize
\end{table}

\begin{table}[htb]
\caption{Results of simultaneous fitting of the XIS/PIN/GSO spectra (model E$^{\dagger}$).}
\label{fite}
\begin{center}
\small
\begin{tabular}{cccccc}
\hline\hline
Parameters & & 2005 & 2009a & 2009b & 2009c \\
\hline
$N_{\rm H}^a$ & ($10^{23}$) &	1.13$\pm$0.01 	&	1.11$\pm$0.01 	&	1.12$\pm$0.02 	&	1.18$\pm$0.01 	\\
$A_{\rm Fe}^b$ & (solar) &	0.73$\pm$0.03 	&	1.01$\pm$0.08 	&	0.91$\pm$0.07 	&	0.72$\pm$0.06 	\\
$I_{\rm pow}^d$ & &	0.147$_{-0.005}^{+0.001}$ 	&	0.178$_{-0.014}^{+0.009}$ 	&	0.153$_{-0.012}^{+0.006}$ 	&	0.141$_{-0.013}^{+0.009}$ 	\\
$N_{\rm H,2}^g$ & ($10^{24}$) &	0.42$\pm$0.07 	&	1.84$\pm$0.26 	&	0.51$\pm$0.08 	&	1.12$\pm$0.24 	\\
$f_{\rm NH2}^g$ & &	0.20$\pm$0.01 	&	0.26$\pm$0.01 	&	0.17$\pm$0.01 	&	0.14$\pm$0.01 	\\
$R^i$ & ($\Omega/2\pi$) &	0.456$\pm$0.002 	&	0.370$\pm$0.002 	&	0.409$\pm$0.003 	&	0.520$\pm$0.003 	\\
$I_{\rm pow2}^j$ & ($10^{-3}$) &	0.0$_{0.0}^{+1.3}$ 	&	26.2$_{-2.1}^{+6.0}$ 	&	20.2$_{-3.0}^{+7.3}$ 	&	31.8$_{-4.4}^{+6.5}$ 	\\
$\chi^2/dof$ & &	 1.41 	&	 1.37 	&	 1.11 	&	 1.23 	\\
($\chi^2,dof$) & &	(1760.5/1245) 	&	(1900.3/1390) 	&	(1505.7/1351) 	&	(1681.6/1367) 	\\
\hline
\multicolumn{6}{p{15cm}}{$\dagger$: Model E is {\tt constant*phabs*( [pcfabs*highecut*zvphabs*powerlaw + phabs*pexmon + phabs*powerlaw] + apec + apec + phabs*bremss + zgauss)} in xspec. Components between parentheses [ ] represent the AGN emission.}\\
\multicolumn{6}{p{15cm}}{$a,b$: Hydrogen column density and Fe abundance of 
 the absorber. The column density is in unit of cm$^{-2}$ for the {\tt zvphabs} model.}\\
\multicolumn{6}{p{15cm}}{$a,b$: Hydrogen column density and Fe abundance of 
 the absorber. The column density is in unit of cm$^{-2}$ for the {\tt zvphabs} model.}\\
\multicolumn{6}{p{15cm}}{$d$: Normalization at 1 keV of the power-law
 model with a fixed photon index 1.9.}\\
\multicolumn{6}{p{15cm}}{$g,h$: Hydrogen column density and covering raction of the absorber for the {\tt pcfabs} model.}\\
\multicolumn{6}{p{15cm}}{$i$: Fraction of the reflection component for the  {\tt pexmon} model.}\\
\multicolumn{6}{p{15cm}}{$j$: Normalization at 1 keV of the additional
 power-law model with a fixed photon index 1.6.}\\
\end{tabular}
\end{center}
\normalsize
\end{table}

\end{document}